\documentclass{article}
\usepackage[utf8]{inputenc}
\pdfoutput=1

\usepackage{newtxtext}
\usepackage[varvw]{newtxmath} 

\usepackage{amsmath}
\usepackage{hyperref}
\usepackage{physics}
\usepackage[
a4paper,
total={160mm,257mm},
left=25mm,
top=20mm]{geometry}
\usepackage{tikz}
\usepackage{authblk}
\usepackage{tabularx}
\usepackage{booktabs} 

\usepackage{mathdots}
\usepackage{cancel}
\usepackage{color}
\usepackage{siunitx}
\usepackage{array}
\usepackage{multirow}
\usepackage{gensymb}
\usepackage{tabularx}
\usepackage{extarrows}
\usepackage{booktabs}
\usetikzlibrary{fadings}
\usetikzlibrary{patterns}
\usetikzlibrary{shadows.blur}
\usetikzlibrary{shapes}

\newcommand{\OO}{\mathrm{O}}

\newcommand{\ee}{\mathrm{e}}
\newcommand{\ii}{\mathrm{i}}
\newcommand{\asf}{\mathsf{a}}
\DeclareMathOperator{\Sp}{Sp}
\DeclareMathOperator{\SO}{SO}
\DeclareMathOperator{\SU}{SU}
\DeclareMathOperator{\OSp}{OSp}
\DeclareMathOperator{\ch}{ch}
\DeclareMathOperator{\sch}{sch}

\numberwithin{equation}{section}

\hypersetup{colorlinks,breaklinks,
            linkcolor = purple,
            citecolor = teal,
            urlcolor = purple,
            }

\title{\bf Orthosymplectic Superinstanton Counting\\ and\\ Brane Dynamics}
\author{Taro Kimura and Yilu Shao}
\affil{\footnotesize \it Institut de Mathématiques de Bourgogne, Université de Bourgogne, CNRS, France}
\date{}

\begin{document}

\maketitle

\begin{abstract}
We extend the study of superinstantons presented in \cite{Kimura:2019msw} to include orthosymplectic supergroup gauge theories, $B_{n_0|n_1}$, $C_n$, and $D_{n_0|n_1}$. We utilize equivariant localization to obtain the LMNS contour integral formula for the instanton partition function, and we investigate the Seiberg--Witten geometries associated with these theories.
We also explore the brane configurations involving positive and negative branes together with O-planes that realize the orthosymplectic supergroup theories.
\end{abstract}

\tableofcontents

\section{Introduction}

Supergroups are symmetries of graded vector spaces with both commuting and anticommuting variables, and is naturally used in studying supersymmetry \cite{DeWitt:2012mdz} as transformation groups on supermanifolds. They are widely used in the study of disordered systems \cite{Efetov:1997fw,Wegner:2016ahw} such as the Parisi-Sourlas supersymmetry \cite{Parisi:1979ka}. Conformal field theories with supergroups as superconformal symmetries also appear frequently in both string theory and condensed matter physics (see \cite{Quella:2013oda} for a review).

Although typically regarded as a global symmetry, supergroups can also be treated as a local gauge symmetry, leading to a \textit{non--unitary} theory with negative energy spectrums and particles that break the spin-statistic theorem. While the perturbation theory of a $\mathrm{U}(n_0|n_1)$ theory cannot be distinguished from those of a $\mathrm{U}(n_0-n_1|0)=\mathrm{U}(n_0-n_1)$ theory \cite{Alvarez-Gaume:1991ozb,Yost:1991ht,Berkovits:1999im,Vafa:2001qf,Dijkgraaf:2016lym}, the nonperturbative dynamics of the former are markedly different due to the presence of the unstable vacua. As a result, proper treatment of the nonperturbative aspects of these theories is crucial. Much progress has been made in understanding their nonperturbative dynamics, ranging from the brane construction using \textit{nagative branes} \cite{Vafa:2001qf,Okuda:2006fb,Dijkgraaf:2016lym,Nekrasov:2017cih,Nekrasov:2018xsb}, non--unitary holography \cite{Vafa:2014iua} and supergroup Chern–Simons theory as the boundary condition of $\mathcal{N}=4$ $S$-duality \cite{Mikhaylov:2014aoa}. It is also discovered that negative branes will lead to dynamical space-time signature change \cite{Dijkgraaf:2016lym} which gives an explanation of the dualities mentioned in \cite{Hull:1998ym} implying possible background-independence of M-theory. In the recent work \cite{Marino:2022rpz,Schiappa:2023ned} the importance of ``negative'' instantons is also mentioned. For an overview of previous studies on supergroup gauge theory and supermatrix model, see a recent review \cite{Kimura:2023iup}.

In this paper, we focus on anti-selfdual Yang--Mills connections in supergroup gauge theories with $B_{n_0|n_1}$, $C_n$, and $D_{n_0|n_1}$ type supergauge groups (or orthosymplectic), which are natural generalizations of the $A_{n_0|n_1}$ type (or unitary) superinstantons studied in previous work \cite{Kimura:2019msw}.

Our main objective is to derive the super-analog of the Losev-Moore-Nekrasov-Shatashvili (LMNS) contour integral formula using equivariant localization. We also investigate the $\widehat{A}_1$ quiver realizations of orthosymplectic supergroup gauge theories and consider the Seiberg--Witten geometries \cite{Seiberg:1994aj,Seiberg:1994rs} associated with these theories. Through instanton calculus and brane dynamics, we have confirmed their consistency. Additionally, we have established the Hanany-Witten brane constructions in type IIA string theory for both supergroup gauge theories and supergroup quiver gauge theories that employ orthosymplectic supergroups as gauge symmetries. To achieve this, we have utilized a combination of negative branes and orientifold planes. We have also explored the gauging process and the one-to-many correspondence \cite{Kimura:2020lmc} associated with non--uniqueness of the simple root decomposition of the super root system. Then we considered the inclusion of ON$^-$ planes to realize supergroup quiver gauge theories of $D$ type.

The paper is organized as follows: In section \ref{sec:review}, we review the ADHM construction as well as the equivariant localization of unitary supergroup gauge theories performed in \cite{Kimura:2019msw}. Then, in section \ref{sec:osp_inst}, we perform equivariant localization and compute the equivariant index of vector multiplets, fundamental hypermultiplets, and bifundamental hypermultiplets. We also provide the exact form of the LMNS contour integral formula. In section \ref{sec:SW}, we review the construction of the Seiberg--Witten curve for ordinary O/Sp gauge theory \cite{Nekrasov:2004vw,Marino:2004cn} and then derive the corresponding Seiberg--Witten geometries. In section \ref{sec:brane_con}, we consider the brane construction. We first review the negative branes in type IIA string theory and then provide the explicit construction for orthosymplectic supergroup gauge theories and confirmed their consistency.

\section{A Brief Review of Superinstantons}
\label{sec:review}
In this section we will review the previous results on supergroup gauge theory, especially the instanton counting method of unitary, denoted $\mathrm{U}(n_0|n_1)$ or $A_{n_0-1|n_1-1}$ supergroup gauge theory in \cite{Kimura:2019msw}.

\subsection{Supergroup Gauge Theory}
\label{sec:Supergroup_Gauge_Theory}
We consider an Yang--Mills theory with gauge supergroup $G=G_0|G_1$ which has the corresponding Lie superalgebra $\mathfrak{g}=\mathfrak{g}_0 \oplus \mathfrak{g}_1=\operatorname{Lie} G$. Let $\mathcal{S}$ be the $\mathsf{d}$-dimensional Riemannian manifold as a base of the $G$-bundle, $A$ be a $\mathfrak{g}$-valued one-form connection $A\in \Omega^1(\mathcal{S},\mathfrak{g})$ and denote the corresponding curvature $F \in \Omega^2(\mathcal{S},\mathfrak{g})$. The $G$-invariant action is 
\begin{equation}
\begin{aligned}
\label{eq:lag}
S_{\mathrm{YM}}[A]& =-\frac{1}{g_{\mathrm{YM}}^2} \int_{\mathcal{S}} \operatorname{str}(F \wedge \star F)\\
& =-\frac{1}{g^2} \int_{\mathcal{S}} \operatorname{tr}_0(F \wedge \star F)-\left(-\frac{1}{g^2}\right) \int_{\mathcal{S}} \operatorname{tr}_1(F \wedge \star F).
\end{aligned}
\end{equation}
with $\star$ the Hodge star operator. 

Evidences discovered about the structure of a supergroup (see sec. \ref{sec:uni_sg} and \ref{sec:osp_sg}) imply that the supergroup gauge theory may possibly be realized in a unphysical parameter regime of the quiver gauge theory $G_0 \times G_1$ with the off-diagonal blocks (fermionic part) transforming under the bifundamental representation. Thus a supergroup gauge theory with $G = G_0|G_1$ can be realized by a cyclic $\widehat{A}_1$ quiver gauge theory:
\begin{align}
\tikzset{every picture/.style={line width=0.75pt}} 
\begin{tikzpicture}[x=0.75pt,y=0.75pt,yscale=-1,xscale=1,baseline=(current bounding box.center)]

\draw   (362.75,140.54) .. controls (362.75,134.46) and (367.68,129.53) .. (373.75,129.53) .. controls (379.83,129.53) and (384.76,134.45) .. (384.76,140.53) .. controls (384.76,146.61) and (379.84,151.53) .. (373.76,151.54) .. controls (367.69,151.54) and (362.76,146.61) .. (362.75,140.54) -- cycle ;
\draw   (493,141.04) .. controls (493,134.96) and (497.93,130.03) .. (504,130.03) .. controls (510.08,130.03) and (515.01,134.95) .. (515.01,141.03) .. controls (515.01,147.11) and (510.09,152.03) .. (504.01,152.04) .. controls (497.94,152.04) and (493.01,147.11) .. (493,141.04) -- cycle ;
\draw    (379.5,131) .. controls (404,111.5) and (475,111.5) .. (498.5,131) ;
\draw    (378.5,150.5) .. controls (403,169.5) and (474,170) .. (497.5,150.5) ;
\draw   (178.25,139.04) .. controls (178.25,132.96) and (183.18,128.03) .. (189.25,128.03) .. controls (195.33,128.03) and (200.26,132.95) .. (200.26,139.03) .. controls (200.26,145.11) and (195.34,150.03) .. (189.26,150.04) .. controls (183.19,150.04) and (178.26,145.11) .. (178.25,139.04) -- cycle ;
\draw    (255,137.5) -- (294.5,137.5)(255,140.5) -- (294.5,140.5) ;
\draw [shift={(301.5,139)}, rotate = 180] [color={rgb, 255:red, 0; green, 0; blue, 0 }  ][line width=0.75]    (10.93,-4.9) .. controls (6.95,-2.3) and (3.31,-0.67) .. (0,0) .. controls (3.31,0.67) and (6.95,2.3) .. (10.93,4.9)   ;
\draw [shift={(248,139)}, rotate = 0] [color={rgb, 255:red, 0; green, 0; blue, 0 }  ][line width=0.75]    (10.93,-4.9) .. controls (6.95,-2.3) and (3.31,-0.67) .. (0,0) .. controls (3.31,0.67) and (6.95,2.3) .. (10.93,4.9)   ;

\draw (521,133.86) node [anchor=north west][inner sep=0.75pt]  [rotate=-359.98]  {$G_{1}$};
\draw (336,131.96) node [anchor=north west][inner sep=0.75pt]  [rotate=-359.98]  {$G_{0}$};
\draw (127,131.9) node [anchor=north west][inner sep=0.75pt]    {$G_{0} |G_{1}$};
\draw (146,163.4) node [anchor=north west][inner sep=0.75pt]    {$g^{2}$};
\draw (335,163.4) node [anchor=north west][inner sep=0.75pt]    {$g^{2}$};
\draw (514,163.4) node [anchor=north west][inner sep=0.75pt]    {$-g^{2}$};

\end{tikzpicture}
\end{align}
with a unphysical coupling constant $g^2_1=-g^2$ \cite{Dijkgraaf:2016lym}, which leads to a non--unitary theory since the energy spectrum is not bounded from below.

\subsection{Equivariant Localization and the LMNS Formula}
Now we turn to the calculation of the instanton partition function of supergroup gauge theories with gauge group $\mathrm{U}(n_0|n_1)$ which correspondes to $A_{n_0-1|n_1-1}$, performed in \cite{Kimura:2019msw}.

\subsubsection{General Notations}
Throughout this paper, the notations are compatible with \cite{Kimura:2020jxl}. Consider an arbitary vector bundle $X$, with its character in terms of Chern roots
\begin{equation}
\operatorname{ch} \mathbf{E}=\sum_{i=1}^{\mathrm{rk} \mathbf{E}} \ee^{x_i}.
\end{equation}
The corresponding index functor can be written as
\begin{equation}
\label{id_fn}
\mathbb{I}[\mathbf{E}]=\prod_{i=1}^{\operatorname{rk} \mathbf{E}}\left[x_i\right]
\end{equation}
where the bracket symbol $[\ldots]$ is defined by
\begin{equation}
[x]= \begin{cases}x & (\text{4d, cohomology}) \\ 1-\ee^{-x} & (\text{5d, K-theory}) \\ \theta\left(\ee^{-x} ; p\right) & (\text{6d, elliptic})\end{cases}
\end{equation}
for 4d $\mathcal{N}=2$ theory on $\mathcal{S}$, 5d $\mathcal{N}=1$ theory on $\mathcal{S} \times S^1$ and 6d $\mathcal{N}=(1,0)$ theory on $\mathcal{S} \times T^2$.

The Adams operations is defined by
\begin{equation}
\operatorname{ch} \mathbf{E}^{[p]}=\sum_{i=1}^{\mathrm{rk} \mathbf{E}} \ee^{px_i}.
\end{equation}

The character of the dual bundle $\mathbf{E}^\vee$ is defined by 
\begin{equation}
\ch \mathbf{E}^{\vee}=\sum_{i=1}^{\mathrm{rk} \mathbf{E}} \ee^{-x_i}.
\end{equation}
In terms of Adams operation we write $\mathbf{E}^{\vee}=\mathbf{E}^{[-1]}$.

We also denote the alternating sum of anti-symmetrizations of the bundle
\begin{equation}
\wedge_y \mathbf{X}=\sum_{i=0}^{\operatorname{rk} \mathbf{X}}(-y)^i \wedge^i \mathbf{X}
\end{equation}
and use the shorthand notation $\wedge \mathbf{X}=\wedge_1 \mathbf{X}$ frequently.

\subsubsection{ADHM Construction}
In a supergroup gauge theory, the anti-selfdual Yang–Mills connection is called the \textit{superinstantons}. The instanton number $k$ is promoted to $\vvmathbb{Z}_2$-graded $(k_0|k_1)$ as well as the gauge group rank. 

For a general quiver gauge theory defined over quiver $\Gamma=(\Gamma_0,\Gamma_1)$, we denote the gauge nodes $i \in \Gamma_0$, and edges (bifundamental hypermultiplets) by $\Gamma_1$. We assign to every node a supergroup $\mathrm{U}(n_0|n_1)$. Consider the $(k_0|k_1)$-instanton moduli space on $\mathcal{S}=\vvmathbb{C}^2$ and we introduce the $\vvmathbb{Z}_2$-graded supervector space 
\begin{equation}
\begin{aligned}
& N=\vvmathbb{C}^{n_0|n_1}=\vvmathbb{C}_0^{n_0} \oplus \vvmathbb{C}_1^{n_1}=: N_0 \oplus N_1 \\
& K=\vvmathbb{C}^{k_0|k_1}=\vvmathbb{C}_0^{k_0} \oplus \vvmathbb{C}_1^{k_1}=: K_0 \oplus K_1
\end{aligned}
\end{equation}
Here we have two instanton numbers $k_{0,1}$ corresponding to $G_{0,1}$ gauge nodes. The total instanton charge is given by $\mathsf{k} = \operatorname{sdim} K = k_0-k_1$ because negative instantons carry negative topological charges. And it is worth noting that because of the effect of these negative-instantons \cite[sec~3.1.4]{Kimura:2019msw} there are infinite contributions at any sector of fixed instanton number. For example the $\mathsf{k}=1$ sector may have $\mathsf{k}=k_0-k_1=1-0=2-1=3-2=...$. So any dualities between supergroup gauge theories can only be seen non-perturbatively.

\paragraph{Framing and Instanton Bundles} We define the graded framing and instanton bundles over the superinstanton moduli space as follows:
\begin{equation}
\mathbf{N}=\left(\mathbf{N}_i\right)_{i \in \Gamma_0}, \quad \mathbf{K}=\left(\mathbf{K}_i\right)_{i \in \Gamma_0}
\end{equation}
with
\begin{equation}
\mathbf{N}_i=\mathbf{N}_i^0 \oplus \mathbf{N}_i^1, \quad \mathbf{K}_i=\mathbf{K}_i^0 \oplus \mathbf{K}_i^1
\end{equation}
which are further decomposed as
\begin{equation}
\mathbf{N}_i^\sigma=\bigoplus_{\alpha=1}^{n_{i, \sigma}} \mathbf{N}_{i, \alpha}^\sigma, \quad \mathbf{K}_i^\sigma=\bigoplus_{\alpha=1}^{n_{i, \sigma}} \mathbf{K}_{i, \alpha}^\sigma
\end{equation}
Their automorphism groups are
\begin{equation}
\mathrm{GL}(N)=\prod_{i \in \Gamma_0} \mathrm{GL}\left(n_{i, 0}|n_{i, 1}\right), \quad \mathrm{GL}(K)=\prod_{i \in \Gamma_0} \mathrm{GL}\left(k_{i, 0}|k_{i, 1}\right)
\end{equation}
Their equivariant characters are given by summing over the Cartan tori respectively
\begin{equation}
\ch_{\mathsf{T}} \mathbf{N}_{i, \alpha}^\sigma=\ee^{\asf_{i, \alpha}^\sigma},\qquad \ch_{\mathsf{T}} \mathbf{K}_{i, \alpha}^\sigma=\ee^{\phi_{i, a}^\sigma}
\end{equation}
In which the parameters are 
\begin{equation}
\begin{aligned}
& \asf_i=\operatorname{diag}\left(\asf_{i, 1}^0, \ldots, \asf_{i, n_{i, 0}}^0, \asf_{i, 1}^1, \ldots, \asf_{i, n_{i, 1}}^1\right) \in \operatorname{Lie} \mathsf{T}_{N_i} \subset \mathfrak{g l}_{N_i}, \\
& \phi_i=\operatorname{diag}\left(\phi_{i, 1}^0, \ldots, \phi_{i, k_{i, 0}}^0, \phi_{i, 1}^1, \ldots, \phi_{i, k_{i, 1}}^1\right) \in \operatorname{Lie} \mathsf{T}_{K_i} \subset \mathfrak{g l}_{K_i}
\end{aligned}
\end{equation}
therefore the supercharacters are
\begin{equation}
\operatorname{sch}_{\mathsf{T}} \mathbf{N}_i=\ch_{\mathsf{T}} \mathbf{N}_i^0-\ch_{\mathsf{T}} \mathbf{N}_i^1, \quad \operatorname{sch}_{\mathsf{T}} \mathbf{K}_i=\ch_{\mathsf{T}} \mathbf{K}_i^0-\ch_{\mathsf{T}} \mathbf{K}_i^1
\end{equation}

\paragraph{Spacetime Bundle} The fiber of the cotangent bundle at the fixed point $o \in \mathcal{S}$ is given by
\begin{equation}
\mathbf{Q}=T_o^{*} \mathcal{S}=\mathbf{Q}_1 \oplus \mathbf{Q}_2
\end{equation}
with the character
\begin{equation}
\ch_{\mathsf{T}} \mathbf{Q}_i=\ee^{\epsilon_i}=q_i \quad(i=1,2) .
\end{equation}
in which $\epsilon_{1,2}$ are the equivariant parameters on $\Omega$-background. We also denote 
\begin{equation}
\wedge \mathbf{Q}_{1,2}=\sum_k(-1)^k \wedge^k \mathbf{Q}_{1,2}, \quad \wedge \mathbf{Q}=\sum_k(-1)^k \wedge^k \mathbf{Q},
\end{equation}
with characters
\begin{equation}
\operatorname{ch}_{\mathsf{T}} \wedge \mathbf{Q}_{1,2}=1-q_{1,2}, \quad \operatorname{ch}_{\mathsf{T}} \wedge \mathbf{Q}=\left(1-q_1\right)\left(1-q_2\right)
\end{equation}
and $\epsilon_{12}=\epsilon_{1}+\epsilon_2$. We also define $\wedge^2 \mathbf{Q}=\operatorname{det} \mathbf{Q} = \mathbf{Q}_1 \mathbf{Q}_2$, with
\begin{equation}
\operatorname{ch}_{\mathsf{T}} \wedge^2 \mathbf{Q}=q_1 q_2=: q
\end{equation}

\paragraph{Observable Bundle} We define the observable bundle by pullback at the fixed point
\begin{equation}
\mathbf{Y}_o=i_o^* \mathbf{Y}_{\mathcal{S}}=\left(\mathbf{N}_i-\wedge \mathbf{Q} \cdot \mathbf{K}_i\right)_{i \in \Gamma_0} .
\end{equation}
It also has a graded structure $\mathbf{Y}_i=\mathbf{Y}_i^{0} \oplus \mathbf{Y}_i^{1}$.

\subsubsection{Equivariant Index Formula}

In analogy to the ordinary $\mathrm{U}(n)$ gauge theory case, the vector multiplet and the hyper- and half-hypermultiplet bundles can be written as
\begin{subequations}
 \begin{align}
\mathbf{V}_i = \frac{\mathbf{Y}_i^\vee \mathbf{Y}_i}{\wedge \mathbf{Q}}
\, ,
\end{align}
\begin{align}
\mathbf{H}_i^\text{f} = - \frac{{\mathbf{M}}_i^\vee\mathbf{Y}_i}{\wedge \mathbf{Q}}
\, , \qquad
\mathbf{H}_i^\text{af} = - \frac{\mathbf{Y}_i^\vee \widetilde{\mathbf{M}}_i}{\wedge \mathbf{Q}} \, .
\end{align}
\end{subequations}
The $\mathbf{M}_i^{\sigma}$s are the matter bundles, whose fibers are the space of virtual zero modes of the Dirac operator in the instanton background. The corresponding characters are
\begin{subequations}
\begin{equation}
\operatorname{ch}_{\mathsf{T}} \mathbf{M}_i^\sigma=\sum_{f=1}^{n_{i, \sigma}^{\mathrm{f}}} \mathrm{e}^{m_{i, f}^\sigma}, \quad \operatorname{ch}_{\mathsf{T}} \widetilde{\mathbf{M}}_i^\sigma=\sum_{f=1}^{n_{i, \sigma}^{\mathrm{af}}} \mathrm{e}^{\widetilde{m}_{i, f}^\sigma}
\end{equation}
\begin{equation}
\operatorname{ch}_{\mathsf{T}} \mathbf{M}_e=\mathrm{e}^{m_e}=: \mu_e
\end{equation}
\end{subequations}

Expanding the above formula gives
\begin{subequations} 
\begin{align}
\sch_\mathsf{T} \mathbf{V}_i^\text{inst}
& = \ch_\mathsf{T} \mathbf{V}_{i,0}^\text{inst} - \ch_\mathsf{T} \mathbf{V}_{i,1}^{\text{inst}}
=
\sum_{\sigma,\sigma' = 0, 1} (-1)^{\sigma + \sigma'} \ch_\mathsf{T} \mathbf{V}_{i,\sigma\sigma'}^{\text{inst}}
\, , \label{eq:schV_inst} \\
\sch_\mathsf{T} \mathbf{H}_e^\text{inst}
& = \ch_\mathsf{T} \mathbf{H}_{e,0}^{\text{inst}} - \ch_\mathsf{T} \mathbf{H}_{e,1}^{\text{inst}}
=
\sum_{\sigma,\sigma' = 0, 1} (-1)^{\sigma + \sigma'} \ch_\mathsf{T} \mathbf{H}_{e,\sigma\sigma'}^{\text{inst}}
\, \\
\sch_\mathsf{T} \mathbf{H}_i^\text{(a)f,inst}
& = \ch_\mathsf{T} \mathbf{H}_{i,0}^\text{(a)f,inst} - \ch_\mathsf{T} \mathbf{H}_{i,1}^\text{(a)f,inst}
=
\sum_{\sigma,\sigma' = 0, 1} (-1)^{\sigma + \sigma'} \ch_\mathsf{T} \mathbf{H}_{i,\sigma\sigma'}^\text{(a)f,inst}
\end{align}
\end{subequations}
where the components are 
\begin{subequations}
\begin{align}
\mathbf{V}_{i,\sigma\sigma'}^\text{inst}
 & =
 - \det \mathbf{Q}^\vee \cdot \mathbf{K}_i^{\sigma\vee} \mathbf{N}^{\sigma'}_i
 - \mathbf{N}^{\sigma\vee}_i \mathbf{K}^{\sigma'}_i
 + \wedge \mathbf{Q}^\vee \cdot \mathbf{K}^{\sigma\vee}_i \mathbf{K}^{\sigma'}_i
 \\
 \mathbf{H}_{i,\sigma\sigma'}^\text{f,inst}
 & = \mathbf{M}^{\sigma\vee}_i \mathbf{K}^{\sigma'}_i
 \\
 \mathbf{H}_{i,\sigma\sigma'}^\text{af,inst}
 & = \det \mathbf{Q^\vee} \cdot \mathbf{K}^{\sigma\vee}_i \widetilde{\mathbf{M}}^{\sigma'}_i
\end{align}
\end{subequations}

By taking the supercharacter of $\mathbf{V}_i^{\text {inst }}$ we get the contour integral formula or the \textit{Losev–Moore–Nekrasov–Shatashvili (LMNS) formula} \cite{Losev:1997bz,Losev:1997tp,Moore:1997dj} of instanton part of the instanton partition function .

\begin{equation}
\label{eq:super_LMNS} 
Z_{n, k}^{\text {inst }}=\frac{1}{k_{0} ! k_{1} !} \frac{\left[-\epsilon_{12}\right]^{k_0+k_1}}{\left[-\epsilon_{1,2}\right]^{k_0+k_1}} \oint_{\mathrm{JK}} \prod_{\substack{\sigma=0,1 \\ a=1, \ldots, k_\sigma}} \frac{d \phi_a^\sigma}{2 \pi \mathrm{i}} \prod_{\sigma, \sigma^{\prime}=0,1} z_{\sigma \sigma^{\prime}}^{\mathrm{vec}} z_{\sigma \sigma^{\prime}}^{\mathrm{f}} z_{\sigma \sigma^{\prime}}^{\mathrm{af}}
\end{equation}
with
\begin{subequations}
\begin{align}
z_{\sigma\sigma'}^\text{vec}
& =
\begin{cases}
\displaystyle
\prod_{a = 1}^{k_\sigma}
P_\sigma(\phi_{a}^\sigma)^{-1}
\widetilde{P}_\sigma(\phi_a^\sigma + \epsilon_{12})^{-1}
\prod_{a \neq b}^{k_\sigma}
\mathscr{S}(\phi_a^\sigma - \phi_b^{\sigma})^{-1}
& (\sigma = \sigma')
\\
\displaystyle
\prod_{a = 1}^{k_{\sigma'}} P_\sigma(\phi_a^{\sigma'})
\prod_{a = 1}^{k_\sigma} \widetilde{P}_{\sigma'}(\phi_a^\sigma + \epsilon_{12})
\prod_{\substack{a = 1,\ldots, k_\sigma \\ b = 1,\ldots, k_{\sigma'}}}
\mathscr{S}(\phi_b^{\sigma'} - \phi_a^{\sigma})
& (\sigma \neq \sigma')  
\end{cases}
\\
z_{\sigma\sigma'}^\text{f}
& =
\begin{cases}
\displaystyle
\prod_{a = 1}^{k_\sigma} P^\text{f}_\sigma(\phi_a^\sigma)
\\
\displaystyle
\prod_{a = 1}^{k_\sigma} P^\text{f}_{\sigma'}(\phi_a^\sigma)^{-1}
\end{cases} 
\qquad
z_{\sigma\sigma'}^\text{af}
=
\begin{cases}
\displaystyle
\prod_{a = 1}^{k_\sigma} \widetilde{P}^\text{af}_{\sigma'}(\phi_a^\sigma + \epsilon_{12})
& (\sigma = \sigma')
\\
\displaystyle
\prod_{a = 1}^{k_\sigma} \widetilde{P}^\text{af}_{\sigma'}(\phi_a^\sigma + \epsilon_{12})^{-1}
& (\sigma \neq \sigma')  
\end{cases}  
\end{align}
\end{subequations}
with the gauge and matter polynomials
\begin{subequations}
\label{eq:poly_U}
\begin{align}
P_\sigma(\phi) = \prod_{\alpha = 1}^{n_\sigma} [\phi - \mathsf{a}_\alpha^\sigma]
\, , \qquad &
\widetilde{P}_\sigma(\phi) = \prod_{\alpha = 1}^{n_\sigma} [- \phi + \mathsf{a}_\alpha^\sigma]
\, , \\
P^\text{f}_\sigma(\phi) = \prod_{\alpha = 1}^{n^\text{f}_\sigma} [\phi - m_f^\sigma]
\, , \qquad &
\widetilde{P}^\text{af}_\sigma(\phi) = \prod_{\alpha = 1}^{n^\text{af}_\sigma} [- \phi + \widetilde{m}_f^\sigma]
\, . 
\end{align}
\end{subequations}
and the $\mathscr{S}$ function
\begin{equation}
\mathscr{S}(\phi)=\frac{\left[\phi-\epsilon_1\right]\left[\phi-\epsilon_2\right]}{[\phi]\left[\phi-\epsilon_{12}\right]}.
\end{equation}
Notice that the contour integral formula~\eqref{eq:super_LMNS} is actually the same as an $\widehat{A}_1$ quiver gauge theory. These results first appears in \cite[eq.~(3.18)]{Kimura:2019msw} and the above is modified from \cite[eq.~(3.4.34)]{Kimura:2020jxl}.

\subsection{The Seiberg--Witten Curve}
For ordinary pure $\mathrm{U}(n)$ super Yang--Mills, the Seiberg--Witten curve is given by
\begin{equation}
y+\frac{\mathfrak{q}}{y}=\operatorname{det}(x-\phi),
\end{equation}
in which $\phi=\operatorname{diag}(\asf_1,\ldots,\asf_n)$ is the adjoint complex scalar field in the $\mathcal{N}=2$ vector multiplet with its vacuum expectation values (or the Coulomb branch parameters) $\asf_i$. $\mathfrak{q}$ is the instanton counting parameter, for definition see \eqref{eq:counting}. For the $\mathrm{U}(n_0|n_1)$ theory, we simply replace the determinant with the superdeterminant
\begin{equation}
y+\frac{\mathfrak{q}}{y}=\operatorname{sdet}(x-\phi)
\end{equation}
with
\begin{equation}
\operatorname{sdet}(x-\phi)=\frac{\prod_{\alpha=1}^{n_0}\left(x-\asf_\alpha^0\right)}{\prod_{\alpha=1}^{n_1}\left(x-\asf_\alpha^1\right)}
\end{equation}
since we can also diagonalize the matrix as
\begin{equation}
\phi=\operatorname{diag}(\phi_0,\phi_1)=\operatorname{diag}(\asf^0_1,\ldots,\asf^0_{n_0},\asf^1_1,\ldots,\asf^1_{n_1})\;.
\end{equation}
Finally we get the corresponding Seiberg--Witten curve
\begin{equation}
\label{eq:swsuper_a}
y+\frac{q}{y}=\frac{\prod_{\alpha=1}^{n_0}\left(x-\asf^0_\alpha\right)}{\prod_{\alpha=1}^{n_1}\left(x-\asf^1_\alpha\right)}.
\end{equation}
This result \cite[eq.~(8.4)]{Dijkgraaf:2016lym} can be derived by brane constructions with negative branes as well as instanton calculus.

\section{Orthosymplectic Superinstantons}
\label{sec:osp_inst}
In this section we begin to discuss the orthosymplectic supergroup gauge theories. We also note that at the level of partition function, the $C_n$ supergroup makes no difference with the $D_{n_0|n_1}$ supergroup with $n_0=1$ and $n_1=n-1$.

\subsection{Total Framing, Instanton and Observable Bundles}
Since orthosymplectic supergroups are ``real'' groups, we define its ``total'' framing and instanton bundles by only taking the real part, i.e.
\begin{equation}
   \mathbf{N}_{i,\mathrm{tot}}=\frac{1}{2}\left(\mathbf{N}_i\oplus\mathbf{N}_i^{\vee}\right),\quad \mathbf{K}_{i,\mathrm{tot}}=\frac{1}{2}\left(\mathbf{K}_i\oplus\mathbf{K}_i^{\vee}\right).
\end{equation}
and are graded
\begin{equation}
\mathbf{N}_{i,\mathrm{tot}}=\mathbf{N}_{i,\mathrm{tot}}^{0}\oplus\mathbf{N}_{i,\mathrm{tot}}^{1},\quad\mathbf{K}_{i,\mathrm{tot}}=\mathbf{K}_{i,\mathrm{tot}}^{0}\oplus\mathbf{K}_{i,\mathrm{tot}}^{1}.
\end{equation}
The parameters are
\begin{equation}
\begin{aligned}
& \mathrm{a}_i=\operatorname{diag}\left(\pm\mathrm{a}_{i, 1}^0, \ldots, \pm\mathrm{a}_{i, n_{i, 0}}^0,(0), \pm\mathrm{a}_{i, 1}^1, \ldots, \pm\mathrm{a}_{i, n_{i, 1}}^1\right) \in \operatorname{Lie} \mathsf{T}_{N_i}. \\
& \phi_i=\operatorname{diag}\left(\pm\phi_{i, 1}^0, \ldots, \pm\phi_{i, k_{i,0}}^0, \pm\phi_{i, 1}^1, \ldots, \pm\phi_{i, k_{i,1}}^1,(0)\right) \in \operatorname{Lie} \mathsf{T}_{K_i}.
\end{aligned}
\end{equation}
So their supercharacters are
\begin{subequations}
\label{eq:char}
\begin{align}
\sch_{\mathsf{T}} \mathbf{N}_{i,\mathrm{tot}} &=\ch_{\mathsf{T}} \mathbf{N}_{i,\mathrm{tot}}^{0}-\ch_{\mathsf{T}} \mathbf{N}_{i,\mathrm{tot}}^{1}=\sum_{\alpha=1}^{n_0}\ee^{\pm \asf^{0}_{\alpha}}+\chi_n-\sum_{\alpha=1}^{n_1}\ee^{\pm \asf^{1}_{\alpha}}\\
\sch_{\mathsf{T}} \mathbf{K}_{i,\mathrm{tot}} &=\ch_{\mathsf{T}} \mathbf{K}_{i,\mathrm{tot}}^{0}-\ch_{\mathsf{T}} \mathbf{K}_{i,\mathrm{tot}}^{1}=\sum_{a=1}^{k_0}\ee^{\pm \phi^{0}_{a}}-\sum_{a=1}^{\xi}\ee^{\pm \phi^{1}_{a}}-\chi_{k}.
\end{align}
\end{subequations}
Here $\chi_n=0,1$ acts as
\begin{equation}
    \begin{aligned}
        \chi_n=1,\ G=\OSp(2n_0+1|n_1) &\leadsto B_{n_0|n_1}\\
        \chi_n=0,\ G=\OSp(2n_0|n_1) &\leadsto D_{n_0|n_1}
    \end{aligned}
\end{equation}
and we denote $k_{1}=2\xi+\chi_{k}$, with $\xi\in\vvmathbb{Z}_{\geq0}$ and $\chi_{k}\in \{0,1\}$ indicates the parity of $k_1$.

The total observable bundle is defined by 
\begin{equation}
\mathbf{Y}_i=\mathbf{N}_{i,\mathrm{tot}}-\wedge \mathbf{Q} \cdot \mathbf{K}_{i,\mathrm{tot}}
\end{equation}
which is graded by
\begin{equation}
\mathbf{Y}_i=\mathbf{Y}_i^0 \oplus \mathbf{Y}_i^1
\end{equation}
as well.

\subsection{Vector Multiplet}
For gauge group $G_i=\OSp(2n_0+\chi_n|n_1)$, contributions from the vector multiplet can be written as 
\begin{equation}
\begin{aligned}
    \mathbf{V}_i &=\frac{1}{2}(\frac{\mathbf{Y}_i^2-\mathbf{Y}_i^{[2]}}{\wedge\mathbf{Q}})\\
    &= \frac{1}{2(\wedge \mathbf{Q})}(\mathbf{N}_{i,\mathrm{tot}}^2-\mathbf{N}_{i,\mathrm{tot}}^{[2]}-2\wedge \mathbf{Q}\cdot \mathbf{N}_{i,\mathrm{tot}}\,\mathbf{K}_{i,\mathrm{tot}}+(\wedge\mathbf{Q})^2\cdot \mathbf{K}_{i,\mathrm{tot}}^2+(\wedge\mathbf{Q})^{[2]}\cdot \mathbf{K}_{i,\mathrm{tot}}^{[2]})
\end{aligned}
\end{equation}
where the perturbative part is
\begin{equation}
\mathbf{V}_i^{\rm pert}=\frac{1}{2\left(\wedge \mathbf{Q}\right)}\left(\mathbf{N}_{i,\mathrm{tot}}^2-\mathbf{N}_{i,\mathrm{tot}}^{[2]}\right)
\end{equation}
and the remaining is the instanton part 
\begin{equation}
\begin{aligned}
\mathbf{V}_i^{\rm inst}&=\frac{1}{2(\wedge \mathbf{Q})}\left[-2(\wedge \mathbf{Q})\cdot \mathbf{N}_{i,\mathrm{tot}}\,\mathbf{K}_{i,\mathrm{tot}}+(\wedge\mathbf{Q})^2\cdot \mathbf{K}_{i,\mathrm{tot}}^2+(\wedge\mathbf{Q})^{[2]}\cdot \mathbf{K}_{i,\mathrm{tot}}^{[2]}\right]\\
&= \sum_{\sigma,\sigma'}\left[-\mathbf{N}_{i,\mathrm{tot}}^{\sigma}\,\mathbf{K}_{i,\mathrm{tot}}^{\sigma'}
+\frac{1}{2}(\wedge\mathbf{Q})\cdot(\mathbf{K}_{i,\mathrm{tot}}^{\sigma}\,\mathbf{K}_{i,\mathrm{tot}}^{\sigma'})
+\frac{1}{2}(1+(\wedge\mathbf{Q}_{1,2}))\cdot(\mathbf{K}_{i,\mathrm{tot}}^{\sigma})^{[2]}\right].
\end{aligned}
\end{equation}

We also remark here that in the case of a quiver gauge theory, the contributions from the vector multiplet of the neighbouring nodes differ with a sign, i.e.
\begin{equation}
\mathbf{V}^{\rm SpO}_j =\frac{1}{2}(\frac{\mathbf{Y}_j^2+\mathbf{Y}_j^{[2]}}{\wedge\mathbf{Q}})
\end{equation}
due to the fact that in order not to violate the global symmetry, the neighbouring node of an orthosymplectic ($\OSp$) gauge node is an SpO gauge node with its bosonic subgroup $\Sp(n) \times \OO(m)$.
This can also be seen from the brane construction in sec. \ref{sec:brane_quiver}. In this case, the parameters are written as \begin{equation}
\begin{aligned}
& \asf_i=\operatorname{diag}\left(\pm\asf_{i, 1}^0, \ldots, \pm\asf_{i, n_{i, 0}}^0, \pm\asf_{i, 1}^1, \ldots, \pm\asf_{i, n_{i, 1}}^1,(0)\right) \in \operatorname{Lie} \mathsf{T}_{N_i}. \\
& \phi_i=\operatorname{diag}\left(\pm\phi_{i, 1}^0, \ldots, \pm\phi_{i, k_{i,0}}^0,(0), \pm\phi_{i, 1}^1, \ldots, \pm\phi_{i, k_{i,1}}^1\right) \in \operatorname{Lie} \mathsf{T}_{K_i}.
\end{aligned}
\end{equation}
In the following sections we will denote OSp gauge nodes by $\Gamma_0^{\circ}$, and SpO gauge nodes by $\Gamma_0^{\bullet}$.

\paragraph{Equivariant Localization}
For elements in $\Gamma_0^{\circ}$ ($\Gamma_0^{\bullet}$ respectively), by taking the index of the super characters we can get the intanton part of the corresponding instanton partition function
\begin{equation}
\label{osp:index}
Z_i^{\text {vec,inst}}=\mathbb{I}\left[\mathbf{V}_i^{\text {inst}}\right].
\end{equation}

\paragraph{Weyl Group Factors}
The LMNS formula for a gauge theory with gauge group $G$ is given as an integral over the maximal Cartan torus of $\mathrm{GL}(K)$, which is a complexification of $G^\vee$: The corresponding ADHM instanton moduli space is given by a quotient by $G^\vee$. Thus apart from the equivariant index derived in the previous section, we need to divide it by the order of the corresponding Weyl group. The Weyl groups and their orders corresponding to classical Lie algebras are shown in the following:
\begin{align}
\begin{tabular*}{.7\textwidth}{@{\extracolsep{\fill}}ccc}
\toprule $G$ & $W(G)$ & $|W(G)|$ \\
\midrule$\mathrm{U}(k)$ & $\mathfrak{S}_{k}$ & $k!$ \\
$\mathrm{SO}(2k+1)$ & $\mathbb{Z}_2^k \rtimes \mathfrak{S}_k$ & $2^k k !$ \\
$\mathrm{Sp}(k)$ & $\mathbb{Z}_2^k \rtimes \mathfrak{S}_k$ & $2^k k !$ \\
$\mathrm{SO}(2k)$ & $H_{k-1} \rtimes \mathfrak{S}_k$ & $2^{k-1} k !$ \\\toprule
\end{tabular*} 
\end{align}
where $H_{k-1}$ is an even subgroup of $\mathbb{Z}_2^k$.

We now consider the LMNS formula for the supergroup gauge theory with only one $B_{n_0|n_1}$ or $D_{n_0|n_1}$ supergroup gauge node. The dual group $G^\vee$ is $\mathrm{SpO}$, $\OSp$ for $G = \OSp$, $\mathrm{SpO}$, respectively. The Weyl group of the supergroup with the maximal bosonic subgroup $G_0 \times G_1$ is given by the product $W(G_0) \times W(G_1)$.
Hence, for the case $G=\OSp$ with instanton number $(k_0|k_1=2\xi+\chi_k)$, the dual group is $G^\vee = \mathrm{SpO}(k_0|k_1)$, and we have
\begin{equation}
\abs{W(\mathrm{SpO}(k_0|k_1))}=2^{k_{0}}k_{0}!2^{\xi-\chi_k}\xi! .
\end{equation}

\paragraph{Contour Integral Formula}
Here we demonstrate the OSp case with single gauge node $\OSp(2n_0+\chi_n|n_1)$ in detail. By expanding the index for the instanton part \eqref{osp:index}, we get: 
\begin{equation}
\label{osp:Z}
\begin{aligned}
Z_{k_{0}|k_{1}}
&=\frac{1}{2^{k_0+\xi-\chi_k}k_0!\xi!}\oint_{\mathrm{JK}}  \prod_{a=1}^{k_{0}}\frac{\dd \phi^{0}_{a}}{2\pi\iota}\,z^{\OO(2n_0+\chi_n)}_{k_{0}}(\phi^0,\asf^0) \prod_{a=1}^{\xi}\frac{\dd \phi^{1}_{a}}{2\pi\iota}\,z^{\Sp(n_{1})}_{k_{1}}(\phi^1,\asf^1)\\
&\quad\times \prod_{a=1}^{k_{0}}\prod_{b=1}^{\xi}\frac{[\pm \phi^{0}_a\pm \phi^{1}_b+\epsilon_{1}][\pm \phi^{0}_a\pm \phi^{1}_b+\epsilon_2]}{[\pm \phi^{0}_a\pm \phi^{1}_b][\pm \phi^{0}_a\pm \phi^{1}_b+\epsilon_{12}]}\\
&\quad\times \prod_{a=1}^{\xi}\prod_{\alpha=1}^{n_{0}}[\pm\phi^{1}_a\pm \asf^{0}_\alpha]
\prod_{a=1}^{k_{0}}\prod_{\alpha=1}^{n_{1}}[\pm\phi^{0}_a\pm \asf^{1}_\alpha]\\
&\quad\times\left(\prod_{\alpha=1}^{n_{0}}[\pm \asf^{0}_{\alpha}]\prod_{a=1}^{k_{0}}\frac{[\pm \phi_{a}^{0} +\epsilon_{1}][\pm \phi_{a}^{0} +\epsilon_2]}{[\pm \phi_{a}^{0}][\pm \phi_{a}^{0} +\epsilon_{12}]}\right)^{\chi_{k}}\\
&\quad\times\left(\prod_{a=1}^{\xi}[\pm \phi_{a}^{1}]\right)^{\chi_n}\times\left([0]\right)^{\chi_n\chi_{k}}.
\end{aligned}
\end{equation}
with $k_1=2\xi+\chi_k$ and the instanton partition function for ordinary $BCD$ groups are
\begin{subequations}
\begin{align}
z^{\OO(2n+\chi_n)}_{k} &= \left(\frac{\left[\epsilon_{12}\right]}{\left[\epsilon_{1,2}\right]}\right)^k  \frac{\prod_{a=1}^k\left[ \pm 2 \phi_a\right]\left[ \pm 2 \phi_a+\epsilon_{12}\right]}{\prod_{\alpha=1}^n\prod_{a}\left[ \pm \phi_a \pm \asf_\alpha\right] \left(\prod_a\left[ \pm \phi_a\right]\right)^{\chi_n}}\cr
&\quad\times\prod_{a<b}\frac{\left[ \pm \phi_a \pm \phi_b\right]\left[ \pm \phi_a \pm \phi_b+\epsilon_{12}\right]}{\left[ \pm \phi_a \pm \phi_b+\epsilon_1\right]\left[ \pm \phi_a \pm \phi_b+\epsilon_2\right]}
\\
z^{\Sp(n)}_{k=2\xi+\chi_k} &= \left(\frac{\left[\epsilon_{12}\right]}{\left[\epsilon_{1,2}\right]}\right)^k \frac{1}{\prod^{\xi}_{\alpha=1}\prod_{a}\left[ \pm \phi_a \pm \asf_\alpha\right] \prod_{a=1}^{\xi}\left[\pm 2 \phi_a+\epsilon_1\right]\left[\pm 2 \phi_a+\epsilon_2\right]}\cr
&\quad\times\left(\frac{\prod_a\left[ \pm \phi_a\right]\left[ \pm \phi_a+\epsilon_{12}\right]}{\left[\epsilon_1\right]\left[\epsilon_2\right]\prod_\alpha\left[\pm \asf_\alpha\right] \prod_{a}\left[\pm \phi_a+\epsilon_1\right]\left[\pm \phi_a+\epsilon_2\right]}\right)^{\chi_k}\cr
&\quad\times\prod_{a<b}\frac{\left[ \pm \phi_a \pm \phi_b \right]\left[ \pm \phi_a \pm \phi_b+\epsilon_{12}\right]}{\left[ \pm \phi_a \pm \phi_b+\epsilon_1\right]\left[ \pm \phi_a \pm \phi_b+\epsilon_2\right]}.
\end{align}
\end{subequations}

Formula \eqref{osp:Z} coincides with the double SO-Sp half-bifundamental formula in \cite[eq.~(A.40)]{Hollands:2010xa} with the mass of the half-hyper multiplets $m_e=0$. Thus the orthosymplectic gauge theory has a realization as an $\widehat{A}_1$ quiver gauge theory:
\begin{align}
\label{fig:osp_sup_quiver}
    \tikzset{every picture/.style={line width=0.75pt}} 
\begin{tikzpicture}[x=0.75pt,y=0.75pt,yscale=-1,xscale=1,baseline=(current bounding box.center)]

\draw   (251.75,52.54) .. controls (251.75,46.46) and (256.68,41.53) .. (262.75,41.53) .. controls (268.83,41.53) and (273.76,46.45) .. (273.76,52.53) .. controls (273.76,58.61) and (268.84,63.53) .. (262.76,63.54) .. controls (256.69,63.54) and (251.76,58.61) .. (251.75,52.54) -- cycle ;
\draw   (382,53.04) .. controls (382,46.96) and (386.93,42.03) .. (393,42.03) .. controls (399.08,42.03) and (404.01,46.95) .. (404.01,53.03) .. controls (404.01,59.11) and (399.09,64.03) .. (393.01,64.04) .. controls (386.94,64.04) and (382.01,59.11) .. (382,53.04) -- cycle ;
\draw    (268.5,43) .. controls (293,23.5) and (364,23.5) .. (387.5,43) ;
\draw    (267.5,62.5) .. controls (292,81.5) and (363,82) .. (386.5,62.5) ;
\draw   (98.25,51.04) .. controls (98.25,44.96) and (103.18,40.03) .. (109.25,40.03) .. controls (115.33,40.03) and (120.26,44.95) .. (120.26,51.03) .. controls (120.26,57.11) and (115.34,62.03) .. (109.26,62.04) .. controls (103.19,62.04) and (98.26,57.11) .. (98.25,51.04) -- cycle ;
\draw    (175,49.5) -- (214.5,49.5)(175,52.5) -- (214.5,52.5) ;
\draw [shift={(221.5,51)}, rotate = 180] [color={rgb, 255:red, 0; green, 0; blue, 0 }  ][line width=0.75]    (10.93,-4.9) .. controls (6.95,-2.3) and (3.31,-0.67) .. (0,0) .. controls (3.31,0.67) and (6.95,2.3) .. (10.93,4.9)   ;
\draw [shift={(168,51)}, rotate = 0] [color={rgb, 255:red, 0; green, 0; blue, 0 }  ][line width=0.75]    (10.93,-4.9) .. controls (6.95,-2.3) and (3.31,-0.67) .. (0,0) .. controls (3.31,0.67) and (6.95,2.3) .. (10.93,4.9)   ;

\draw (373,83.86) node [anchor=north west][inner sep=0.75pt]  [rotate=-359.98]  {$\mathrm{Sp}(n_{1})$};
\draw (236,83.96) node [anchor=north west][inner sep=0.75pt]  [rotate=-359.98]  {$\mathrm{SO}(2n_{0}+\chi_n)$};
\draw (69,81.9) node [anchor=north west][inner sep=0.75pt]    {$\mathrm{OSp}( 2n_{0}+\chi_n |n_{1})$};

\end{tikzpicture}
\end{align}

We note again that when we set $\chi_{n,k}=1$, it happens that due to the fermonic zero modes in the ADHM supersymmetric quantum mechanics the whole partition function becomes zero since the bifundamental half-hypermultiplets are massless.

Although the index functor \label{id_fn} works for $4/5/6$d, in the case 5d $\mathcal{N}=1$ theory there could be a discrete $\theta$-angle due to $\pi_4(\mathrm{Sp}(N)) = \mathbb{Z}_2$ as analogy to ordinary 5d $\Sp(n)$ gauge theory. And the $k$-instanton moduli space have contributions from two connected components of the $\mathrm{O}(k)_\pm$ group. The integral formula may captures only a specific subset of instanton partition functions in 5d, which is left for future work\footnote{We thank the anonymous referee for pointing this out.}.

\subsection{Fundamental and Bifundamental (Half-)Hypermultiplets}
\label{sec:hyper}
More generally, for a quiver gauge theory corresponding to quiver $\Gamma=(\Gamma_0, \Gamma_1)$ with ortheosymplectic gauge nodes, we can have (anti-)fundamental hypermultiplets and bifundamental half-hypermultiplets \cite{Hollands:2010xa,Zhang:2019msw}.
\subsubsection{Fundamental Hypermultiplets}
In order to be consistent with global symmetries, the (anti-)fundamental hypermultiplet associated to an orthosymplectic ($\OSp$) gauge node should have an SpO type flavor symmetry. For a(n) (anti-)fundamental hypermultiplet with flavor symmetry $\text{SpO}(n^{\mathrm{f},0}_i|2n^{\mathrm{f},1}_i+\chi_f)$ connected to the $i$-th node, we assign the total matter bundles \cite{Kimura:2019msw,Kimura:2020jxl} by
\begin{equation}
\left(\mathbf{M}_{\mathrm{tot},i}\right)_{i \in \Gamma_0}=\frac{1}{2}\left(\mathbf{M}_i+\mathbf{M}_i^{\vee}\right), \quad
\left(\widetilde{\mathbf{M}}_{\mathrm{tot},i}\right)_{i \in \Gamma_0}=\frac{1}{2}\left(\widetilde{\mathbf{M}}_i+\widetilde{\mathbf{M}}_i^{\vee}\right)
\end{equation}
with graded mass parameter $m_{i,f}^{\sigma}$ or $\tilde{m}^{\sigma}_{i,f}$, and $\chi_\mathrm{(a)f}=0,1$ depending on the parity. The supercharacters associated with the flavor symmetry are
\begin{subequations}
\begin{align}
\operatorname{sch}_{\mathsf{T}}\mathbf{M}_{\mathrm{tot},i}&= \operatorname{ch}_{\mathsf{T}}\mathbf{M}_{\mathrm{tot},i}^0-\operatorname{ch}_{\mathsf{T}}\mathbf{M}_{\mathrm{tot},i}^1 = \sum_{f=1}^{n_{i, 0}^{\mathrm{f}}} \mathrm{e}^{\pm m_{i, f}^0}-\sum_{f=1}^{n_{i, 1}^{\mathrm{f}}} \mathrm{e}^{\pm m_{i, f}^1}-\chi_\mathrm{f}, \\
\operatorname{sch}_{\mathsf{T}}\widetilde{\mathbf{M}}_{\mathrm{tot},i}&=\operatorname{ch}_{\mathsf{T}}\widetilde{\mathbf{M}}_{\mathrm{tot},i}^0-\operatorname{ch}_{\mathsf{T}}\widetilde{\mathbf{M}}_{\mathrm{tot},i}^1 = \sum_{f=1}^{n_{i, 0}^{\mathrm{af}}} \mathrm{e}^{\pm \tilde{m}_{i, f}^0 }-\sum_{f=1}^{n_{i, 1}^{\mathrm{af}}} \mathrm{e}^{\pm \tilde{m}_{i, f}^1}-\chi_\mathrm{af}.
\end{align}   
\end{subequations}

Then the (anti-)fundamental hypermultiplet contributions to the (extended) tangent bundle of the instanton moduli space can be written as
\begin{equation}
\label{eq:ospf}
\mathbf{H}_i^\text{f} = - \frac{{\mathbf{M}}_{\mathrm{tot},i}\,\mathbf{Y}_i}{\wedge \mathbf{Q}}:=\mathbf{H}_i^\text{f,pert}+\mathbf{H}_i^\text{f,inst}, \quad
\mathbf{H}_i^\text{af} = - \frac{\mathbf{Y}_i\,\widetilde{\mathbf{M}}_{\mathrm{tot},i}}{\wedge \mathbf{Q}}:=\mathbf{H}_i^\text{af,pert}+\mathbf{H}_i^\text{af,pert}
\end{equation}
By expanding \eqref{eq:ospf} we can extract the perturbative and  instanton contributions
\begin{subequations}
\begin{align}
\mathbf{H}_i^\text{f,pert} &= -\frac{{\mathbf{M}}_{\mathrm{tot},i}\,\mathbf{N}_{{\rm tot},i}}{\wedge \mathbf{Q}}, \hspace{1.35em} \mathbf{H}_i^\text{f,inst} = {\mathbf{M}}_{\mathrm{tot},i}\,{\mathbf{K}}_{\mathrm{tot},i}\\
\mathbf{H}_i^\text{af,pert} &= -\frac{\mathbf{N}_{{\rm tot},i}\,{\widetilde{\mathbf{M}}}_{\mathrm{tot},i}}{\wedge \mathbf{Q}},\quad \mathbf{H}_i^\text{af,inst} = {\mathbf{K}}_{\mathrm{tot},i}\,\widetilde{\mathbf{M}}_{\mathrm{tot},i}.
\end{align}
\end{subequations}
Recall \eqref{eq:char}, we can now write the contributions of the fundamental hypermultiplets to the instanton partition function as
\begin{equation}
\begin{aligned}
Z_i^{\text {f,inst}}&=\mathbb{I}\left[\mathbf{H}_i^\text{f,inst}\right]\\
&=\prod_{a=1}^{k_{i,0}}\frac{
\prod_{f=1}^{n_{i, 0}^{\mathrm{f}}}\left[\pm \phi_{i,a}^0 \pm m_{i, f}^0\right]}{\prod_{f=1}^{n_{i, 1}^{\mathrm{f}}}\left[\pm \phi_{i,a}^0 \pm m_{i, f}^1\right]\left([\pm \phi_{i,a}^0]\right)^{\chi_{\mathrm{f}}}}\\
&\times \prod_{a=1}^{k_{i,1}}\frac{\prod_{f=1}^{n_{i, 1}^{\mathrm{f}}}\left[\pm \phi_{i,a}^1 \pm m_{i, f}^1\right]\left([\pm m_{i,f}^1]\right)^{\chi_k}([0])^{\chi_{\mathrm{f}}\chi_k}}{\prod_{f=1}^{n_{i, 0}^{\mathrm{f}}}\left[\pm \phi_{i,a}^1 \pm m_{i, f}^0\right]\left([\pm \phi_{i,a}^1]\right)^{\chi_{\mathrm{f}}}\left([\pm m_{i,f}^0]\right)^{\chi_k}},
\end{aligned}    
\end{equation}
and
\begin{equation}
\begin{aligned}
Z_i^{\text {af,inst}}&=\mathbb{I}\left[\mathbf{H}_i^\text{af,inst}\right]\\
&=\prod_{a=1}^{k_{i,0}}\frac{
\prod_{f=1}^{n_{i, 0}^{\mathrm{af}}}\left[\pm \phi_{i,a}^0 \pm m_{i, f}^0\right]}{\prod_{f=1}^{n_{i, 1}^{\mathrm{af}}}\left[\pm \phi_{i,a}^0 \pm m_{i, f}^1\right]\left([\pm \phi_{i,a}^0]\right)^{\chi_{\mathrm{af}}}}\\
&\times \prod_{a=1}^{k_{i,1}}\frac{\prod_{f=1}^{n_{i, 1}^{\mathrm{af}}}\left[\pm \phi_{i,a}^1 \pm m_{i, f}^1\right]\left([\pm m_{i,f}^1]\right)^{\chi_k}([0])^{\chi_{\mathrm{af}}\chi_k}}{\prod_{f=1}^{n_{i, 0}^{\mathrm{af}}}\left[\pm \phi_{i,a}^1 \pm m_{i, f}^0\right]\left([\pm \phi_{i,a}^1]\right)^{\chi_{\mathrm{af}}}\left([\pm m_{i,f}^0]\right)^{\chi_k}},.
\end{aligned}
\end{equation}
which are the superanalogs of \cite[eq.~(42)]{Zhang:2019msw}. Here we found that when we set $\chi_{\mathrm{(a)f},k}=1$ the instanton partition function vanishes due to the fermonic zero modes. Similiar phenomenon can be seen if we consider a symplectic gauge node and an orthogonal flavor node, and take one of the mass parameters to zero (corresponds to $\chi_{\mathrm{(a)f}}=1$).

We can also define the supermatter polynomials similiar to \eqref{eq:poly_U} by
\begin{equation}
P_{i,\sigma}^{\mathrm{f}}(\phi)=\prod_{f=1}^{n_{i,\sigma}^{\mathrm{f}}}\left[\pm\phi\pm m_{i,f}^\sigma\right], \quad \widetilde{P}_{i,\sigma}^{\mathrm{af}}(\phi)=\prod_{f=1}^{n_{i,\sigma}^{\mathrm{af}}}\left[\pm \phi\pm\widetilde{m}_{i,f}^\sigma\right]
\end{equation}
So in terms of supermatter polynomials we can write the fundamental hypermultiplets as
\begin{subequations}
\begin{align}
Z_i^{\text {f,inst}}&=\prod_{\sigma,\sigma'}\left\{P_{i,\sigma'}^{\mathrm{f}}(\phi_{i,a}^\sigma)\right\}^{\Pi(\sigma,\sigma')}\\
Z_i^{\text {af,inst}}&=\prod_{\sigma,\sigma'}\left\{\widetilde{P}_{i,\sigma'}^{\mathrm{af}}(\phi_{i,a}^\sigma)\right\}^{\Pi(\sigma,\sigma')}.
\end{align} 
\end{subequations}
with $\Pi(\sigma,\sigma')=(-1)^{-(\sigma+\sigma')}$.

\subsubsection{Bifundamental Half-hypermultiplets}
The contributions from massless bifundamental half-hypermultiplet between $i$-th and $j$-th nodes, i.e. the edge $e:i \to j$ is \cite[eq.~(A.39)]{Hollands:2010xa}:
\begin{equation}
    \mathbf{H}_{e:i \to j}=\frac{1}{2}(\frac{\mathbf{Y}_i^{i} \mathbf{Y}_i^{j}}{\wedge\mathbf{Q}}):=\mathbf{H}_e^{\text{pert}}+\mathbf{H}_e^{\text{inst}}.
\end{equation}
with
\begin{equation}
\begin{aligned}
\mathbf{H}^{\text {pert }}_e &=\frac{\mathbf{N}_{\mathrm{tot},i}\mathbf{N}_{\mathrm{tot},j}}{2(\wedge\mathbf{Q})}\\
\mathbf{H}^{\text {inst }}_e &=\frac{1}{2}\left[-\mathbf{N}_{\mathrm{tot},i}\mathbf{K}_{\mathrm{tot},j}-\mathbf{K}_{\mathrm{tot},i}\mathbf{N}_{\mathrm{tot},j}+(\wedge\mathbf{Q})\cdot\mathbf{K}_{\mathrm{tot},i}\mathbf{K}_{\mathrm{tot},j}\right]\\
\end{aligned}
\end{equation}
The instanton contributions are thus obtained by considering the square root of a full bifundamental hypermultiplet. In terms of the supergauge polynomial and $\mathscr{S}$ function, the bifundamental half-hypermultiplet contribution is 
\begin{equation}
\begin{aligned}
Z_e^{\text{bf,inst}}&=\mathbb{I}\left[\mathbf{H}_{e}^\text{inst}\right]\\
&=\prod_{\sigma,\sigma'}\left\{\prod_{\alpha_i,a_j}P_{i,\sigma'}(\phi^{\sigma}_{j,a_j})\prod_{\alpha_j,a_i}P_{j,\sigma'}(\phi^{\sigma}_{i,a_i})\prod_{a_i,b_j}\mathscr{S}(\pm\phi^{\sigma}_{a_i}\pm\phi^{\sigma'}_{b_j})^{-1}\right\}^{\Pi(\sigma,\sigma')}.
\end{aligned}
\end{equation}
This is a superanalog of \cite[eq.~(45)]{Zhang:2019msw}.

\subsection{Quiver Gauge Theory}
\label{sec:inst_quiver}
Let $\Gamma(\Gamma_0,\Gamma_1)$ be a quiver with gauge nodes $G_i$ of orthosymplectic supergroups. We can formally write down the instanton partition function by combining the ingredients from section \ref{sec:osp_inst}. The instanton partition function for instanton number $(k_0|k_1)$ reads:
\begin{equation}
Z^{\text{inst.}}_{(k_0|k_1)}=\prod_{i\in \Gamma_0}\frac{1}{\left|W_i\right|} \oint_{\mathsf{T}_K} \prod_{a,\sigma} \frac{d \phi^{\sigma}_{i, a}}{2 \pi\ii}Z_i^{\text{vec,inst.}}Z_i^{\text{f,inst.}}Z_i^{\text{af,inst.}}\left(\prod_{e\in \Gamma_1}Z^{\text{bf,inst.}}_e\right).
\end{equation}
where $\left|W_i\right|$ is the order of the Weyl group of $G^\vee_i$. For the $\OSp$ case ($i\in \Gamma_0^{\circ}$) with instanton number $(k_{i,0}|k_{i,1}=2\xi_i+\chi_{i,k})$ we have $G_i^\vee=\mathrm{SpO}(k_{i,0}|k_{i,1})$ and 
\begin{equation}
\abs{W_i(\mathrm{SpO}(k_{i,0}|k_{i,1}))}=2^{k_{i,0}}(k_{i,0})!2^{\xi_i}\xi_i!
\end{equation}
For the SpO case ($i\in \Gamma_0^{\bullet}$) with instanton number $(k_{i,0}=2\zeta_i+\chi_{i,k}|k_{i,1})$ we have $G_i^\vee=\OSp(k_{i,0}|k_{i,1})$ and
\begin{equation}
\abs{W_i(\OSp(k_{i,0}|k_{i,1}))}=2^{\zeta_i}\zeta_i!2^{k_{i,1}}(k_{i,1})!
\end{equation}
The gauge nodes of the quiver theory should be alternating OSp and SpO nodes, as we have claimed before. This can also be seen from the brane constructions in sec. \ref{sec:brane_quiver}.

\section{Seiberg--Witten Geometry}
\label{sec:SW}
\subsection{The $BCD$ Group Case}
The Seiberg--Witten curve for ordinary $BCD$-type gauge theory was given in \cite{Danielsson:1995is,Brandhuber:1995zp,Martinec:1995by}. An alternative approach is applying the instanton counting method \cite{Nekrasov:2004vw,Marino:2004cn}. They consider an embedding of $BCD$-type theory into an $A$-type theory and derived corresponding Seiberg--Witten curve with certain parameters. Here we provide an alternative method that can be generalized for supergroup gauge theories. In this section we only consider 4d theories. For 5d or 6d theories modification are needed.
\subsubsection{The Case of $B_n$ and $D_n$}
Consider the $\SO(2n+\chi)$ theory. It can be regarded as an 4d $\SU(2n)$ theory with $4-2\chi$ massless fundamental matter multiplets , and imposing certain vacuum expectation value of the scalar field $\phi$, namely
\begin{equation}
\langle\phi\rangle=\operatorname{diag}\left\{\ii \asf_1,-\ii \asf_1, \ldots, \ii \asf_n,-\ii \asf_n\right\}
\end{equation}
The Seiberg--Witten curve for $\SO(2n+\chi)$ theory is then
\begin{equation}
\label{eq:sw_bd}
y+\frac{\Lambda^{4 n+2 \chi-4} x^{4-2 \chi}}{y}=\prod_{l=1}^n\left(x^2-\asf_l^2\right)
\end{equation}

\begin{figure}[htbp]
	\centering
	\includegraphics{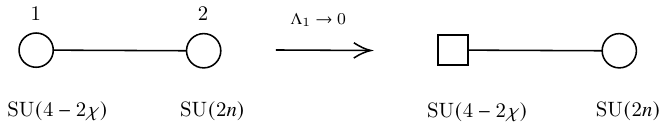}
	\caption{$B_n$ and $D_n$ as $A_2$ quiver gauge theory.}
	\label{fig:bd}
\end{figure}

This result can also be derived from the decoupling limit of an $A_2$ quiver gauge theory. Consider the quiver in fig. \ref{fig:bd}, the two gauge nodes are $\SU(4-2\chi)$ and $\SU(2n)$ respectively. Recall the fundamental characters \cite{Nekrasov:2012xe} (with notations in \cite[4.3.6]{Kimura:2020jxl}) are
\begin{equation}
\begin{aligned}
\label{eq:bd}
\chi_1\left(y_{1,2}\right)&:y_1+\frac{\Lambda^{b_1}_1 y_2}{y_1}+\frac{\Lambda^{b_1}_1\Lambda^{b_2}_2}{y_2}&=\mathsf{T}_1(x), \\
\chi_2\left(y_{1,2}\right)&:y_2+\frac{\Lambda^{b_2}_2 y_1}{y_2}+\frac{\Lambda^{b_1}_1 \Lambda^{b_2}_2}{y_1}&=\mathsf{T}_2(x),
\end{aligned}
\end{equation}
with 
\begin{equation}
    b_i=2n_i-\sum_{j}n_j
\end{equation}
where $n_i$ is the rank of the $i$-th gauge node, $j$ denotes the neighbouring nodes of $i$. In our case,
\begin{equation}
    b_1=8-4\chi-2n,\quad b_2=4n+2\chi-4.
\end{equation}
For node 1, the Coulomb branch parameters in gauge polynomial is replaced by the fundamental mass parameter $m_f$. Since here all fundamental matter multiplets are massless, we have
\begin{equation}
    \mathsf{T}_1(x)=\prod_{f=1}^{n_1}\left(x-m_f\right)=x^{4-2\chi}.
\end{equation}
And the gauge polynomial is
\begin{equation}
    \mathsf{T}_2(x)=\prod_{l=1}^n\left(x^2-\asf_l^2\right).
\end{equation}
Substitute into \eqref{eq:bd} and taking the decoupling limit $\Lambda_1\to 0$, we have
\begin{equation}
\begin{aligned}
y_1&=x^{4-2\chi}, \\
y_2+\frac{\Lambda^{4n+2\chi-4}_2 y_1}{y_2}&=\prod_{l=1}^n\left(x^2-\asf_l^2\right).
\end{aligned}
\end{equation}
Eliminate $y_1$ and rename $y_2\to y$, $\Lambda_2\to \Lambda$, we recover \eqref{eq:sw_bd}.

\subsubsection{The Case of $C_n$}
The $\Sp(n)$ theory can be regarded as an $\SU(2n+2)$ theory, with certain vacuum expectation value of the scalar field $\phi$
\begin{equation}
\langle\phi\rangle=\operatorname{diag}\left\{\ii \asf_1,-\ii \asf_1, \ldots, \ii \asf_n,-\ii \asf_n, 0,0\right\}
\end{equation}
so the curve is given by
\begin{equation}
\label{eq:sw_c}
y+\frac{\Lambda^{4 n+4}}{y}=x^2 \prod_{l=1}^n\left(x^2-\asf_l^2\right)+2 \Lambda^{2 n+2}.
\end{equation}
The derivation of the extra $2 \Lambda^{2 n+2}$ factor can be found in \cite{Nekrasov:2004vw,Marino:2004cn}.

\begin{figure}[htbp]
	\centering
	\includegraphics{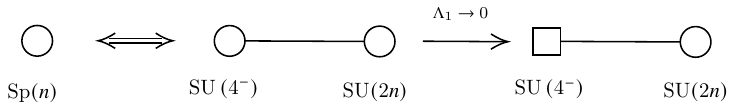}
	\caption{$C_n$ as $A_2$ quiver gauge theory with supergroup gauge nodes.}
	\label{fig:c}
\end{figure}

We can also regarded it as an $A_2$ quiver gauge theory, but this time a little bit different. The two gauge nodes are $\SU(2n)$ and $\SU(4^-)=\mathrm{SU}(0|4)$ respectively \cite{Nekrasov:2004vw}. We can think of this as supergroup gauge node in the following way.

For an $\SU(n)$ gauge theory with $n^f$ fundamental matter multiplets, its Seiberg--Witten curve can be written as
\begin{equation}
\label{eq:sw_a}
    y+\frac{P(x)\Lambda^{2n-n^{\mathrm{f}}}}{y}=\prod_{\alpha=1}^n (x-\asf_{\alpha})
\end{equation}
with the matter polynomial
\begin{equation}
P(x)=\prod_{f=1}^{n^{\mathrm{f}}}\left(x-m_f\right) .
\end{equation}
Now we perform a change of variable $y\to y\sqrt{P(x)}$, the curve \eqref{eq:sw_a} becomes
\begin{equation}
y+\frac{\Lambda^{2n-n^{\mathrm{f}}}}{y}=\frac{\prod_{\alpha=1}^n (x-\asf_{\alpha})}{\sqrt{P(x)}}
\end{equation}
which has similar form as \eqref{eq:swsuper_a}. The Seiberg--Witten curve of a $\mathrm{SU}(n_0|n_1)$ supergroup gauge theory is the same as an $\SU(n_0)$ SQCD (super quantum chromodynamics) with $2n_1$ flavors with pairwisely identified Coulomb branch parameters \cite{Dijkgraaf:2016lym}. Thus an $\mathrm{SU}(n^{\mathrm{f}}_0|n^{\mathrm{f}}_1)$ flavor group will generate a matter polynomial of the form
\begin{equation}
P(x)=\frac{\prod_{f=1}^{n^{\mathrm{f}}_0}\left(x-m_f\right)}{\prod_{f=1}^{n^{\mathrm{f}}_1}\left(x-m_f\right)}.
\end{equation}
For the superflavor group $\mathrm{SU}(n^{\mathrm{f}}_0|n^{\mathrm{f}}_1)$ we also define $n^{\mathrm{f}}=n^{\mathrm{f}}_0-n^{\mathrm{f}}_1$.

In our case, we can think of the $C_n$ gauge theory as gauging a $\SU(2n)$ gauge theory with massless $\SU(4^-)$ flavors. The $\SU(4^-)$ node gives the contribution
\begin{equation}
    \sqrt{P(x)}=\frac{1}{\left(\prod_{f=1}^4 (x-m_f)\right)^{1/2}}=\frac{1}{\sqrt{x^4}}=\frac{1}{x^2}
\end{equation}
combining with the curve of $\SU(2n)$ theory and rename the variables we get
\begin{equation}
    y+\frac{\Lambda^{4n+4}}{y}=\frac{\prod_{l=1}^n\left(x^2-\asf_l^2\right)+2/x^2}{1/x^2}=x^2\prod_{l=1}^n\left(x^2-\asf_l^2\right)+2
\end{equation}
which is exactly \eqref{eq:sw_c}. Note that due to the domain change in \cite[appendix.~A]{Nekrasov:2004vw} (see \cite{Marino:2004cn} also) there is an extra $2/x^2$ term.

\subsection{The $BCD$ Supergroup Case}
We use the same strategy. For an $B_{n_0|n_1}$ or $D_{n_0|n_1}$ (corresponds to $\OSp(2n_0+\chi_n|n_1)$ theories respectively) theory, we can embed it into a quiver with unitary gauge nodes. 

\subsubsection{$\widehat{A}_1$ Quiver Gauge Theory}
\begin{figure}[htbp]
	\centering
	\includegraphics{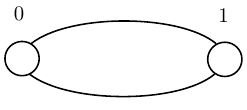}
	\caption{An $\widehat{A}_1$ quiver.}
	\label{fig:a1}
\end{figure}
It is stated in \cite{Nekrasov:2018xsb} that for $G_{\Gamma}$ a simple Lie group with the corresponding Dynkin diagram given by the quiver $\Gamma=A D E$, then the Seiberg-Witten curve is constructed by characters of the fundamental representations of $G_{\Gamma}$ (or affine characters for $\Gamma=\widehat{A D E}$ ). 

Here based on the observation of \eqref{fig:osp_sup_quiver} we start with an $\widehat{A}_1$ affine quiver. The affine fundamental character of this affine quiver is given according to the Cartan matrix
\begin{equation}
C=\left(\begin{array}{cc}
2 & -2 \\
-2 & 2
\end{array}\right).
\end{equation}
And the rule is that we generate the fundamental character by iWeyl reflection \cite{Nekrasov:2012xe}. That is, starting with $y_i$, and continuously replacing the numerator by the following rule
\begin{equation}
\label{eq:affine}
\begin{aligned}
y_0 &\rightarrow y_0 \cdot \frac{y_1^2}{y_0^2} \mathfrak{q}_0 = \frac{y_1^2}{y_0}\mathfrak{q}_0\\
y_1 &\rightarrow y_1 \cdot \frac{y_0^2}{y_1^2} \mathfrak{q}_1 = \frac{y_0^2}{y_1}\mathfrak{q}_1
\end{aligned}
\end{equation}
where $\mathfrak{q}_i$s are the instanton counting parameters, which are multiples of the coupling constants
\begin{equation}
\label{eq:counting}
    \mathfrak{q}_i=\lambda_i \Lambda_i^{2n_i-2\sum_j n_j} 
\end{equation}
with $i,j$ neighboring nodes. For the superconformal case, $b_i = 0$. The coupling constant becomes a dimensionless parameter $\lambda_i$.

In the case of $\widehat{A}_1$ quiver, we have
\begin{equation}
    y_0+\frac{y_1^2}{y_0}\mathfrak{q}_0+...
\end{equation}
which goes on infinitely. 

\subsubsection{Supergroup Case}\label{sec:A1^_supergroup} 
However, the series is finite if we are considering the $\widehat{A}_1$ quiver gauge associated to a supergroup gauge theory. As is seen in \eqref{eq:lag}, the coupling constant satisfies $\tau_1=-\tau_0$, or in exponential form, $\mathfrak{q}_1=\mathfrak{q}_0^{-1}$. We notice that under the rule \eqref{eq:affine}, the following
\begin{equation}
\begin{aligned}
\frac{y_0}{y_1} &\rightarrow \frac{y_0}{y_1} \cdot \frac{y_1^2}{y_0^2} \mathfrak{q}_0 = \frac{y_1}{y_0} \mathfrak{q}_0\\
\frac{y_1}{y_0} &\rightarrow \frac{y_1}{y_0} \cdot \frac{y_0^2}{y_1^2} \mathfrak{q}_1 = \frac{y_0}{y_1} \mathfrak{q}_0^{-1}
\end{aligned}
\end{equation}
forms an involution. The affine character is
\begin{equation}
\chi(y_{0,1}):\ \frac{y_0}{y_1}+\frac{y_1 \mathfrak{q}_0}{y_0}.
\end{equation}
Thus the Seiberg--Witten curve for a general supergroup gauge theory takes the form
\begin{equation}
\label{eq:swsuper_gen}
\frac{y_0}{y_1}+\frac{y_1 \mathfrak{q}_0}{y_0}=\frac{\mathsf{T}_0(x)}{\mathsf{T}_1(x)}.
\end{equation}
with $\mathsf{T}_i$ the gauge polynomials. Let $y=y_0/y_1$ and $\mathfrak{q}=\mathfrak{q}_0$, we have the more familiar expression
\begin{equation}
\label{eq:sw_sggen}
y+\frac{\mathfrak{q}}{y}=\frac{\mathsf{T}_0(x)}{\mathsf{T}_1(x)}.
\end{equation}
This result is rigorously derived from instanton calculus in \cite[eq.~7.81]{Nekrasov:2012xe} including also the case when $\mathfrak{q}_0 \neq \mathfrak{q}_1$. 

\subsubsection{$\widehat{A}_1$ Quiver with Matters}
\begin{figure}[htbp]
	\centering
	\includegraphics{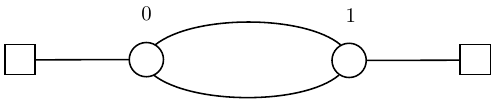}
	\caption{An $\widehat{A}_1$ quiver with flavor nodes.}
	\label{fig:a1f}
\end{figure}

In order to utilize the technique in \cite{Marino:2004cn,Nekrasov:2004vw} to embed an orthosymplectic supergroup gauge theory into a unitary theory, we need to consider adding matter to the $\widehat{A}_1$ quiver which is an asymptotic non--free theory. In general this will leads to a divergent series, but we can still write down the affine characters formally. In this case we change the substitution rules as follows by adding matter polynomials $P_i(x)$, which corresponds to the flavor nodes connected to the gauge nodes.

The iWeyl reflections are now
\begin{equation}
\begin{aligned}
y_0 &\rightarrow \frac{y_1^2}{y_0}\mathfrak{q}_0 P_0(x)\\
y_1 &\rightarrow \frac{y_0^2}{y_1}\mathfrak{q}_1 P_1(x).
\end{aligned}
\end{equation}

\subsection{Seiberg--Witten Curve for $D_{n_0|n_1}$}
\label{sec:sw_Dsg}
\begin{figure}[htbp]
	\centering
	\includegraphics{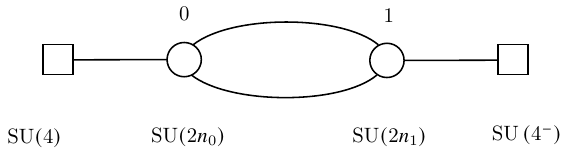}
	\caption{The embedding of $D_{n_0|n_1}$ supergroup gauge theory into a unitary quiver.}
	\label{fig:a1d}
\end{figure}
To manifest the supergroup condition, the matter polynomials need to satisfy a supergroup condition, physically speaking even this theory is realized as a quiver gauge theory, it is still instrinctly a supergroup gauge theory. Thus the fundamental character should remain finite. This implies the \textit{superflavor symmetry}:
\begin{equation}
    P_1(x)=P_0(x)^{-1}.
\end{equation}
Then the supergroup condition is enlarged to $\mathfrak{q}_0 \mathfrak{q}_1 = P_0(x) P_1(x) = 1$. With these condition satisfied, we saw that $P_{0,1}(x)$ cancels each other and the Seiberg--Witten curve remains in the form of \eqref{eq:sw_sggen}.

From fig. \ref{fig:a1d} we see that
\begin{equation}
\begin{aligned}
    \mathsf{T}_0 &= \prod_{\alpha=1}^{n_0}(x^2-(\asf^0_\alpha)^2)\\
    \mathsf{T}_1 &= \prod_{\alpha=1}^{n_1}(x^2-(\asf^1_\alpha)^2)+\frac{2}{x^2}.
\end{aligned}
\end{equation}
Here we take $P_0(x)=x^2$ with
\begin{equation}
\langle\phi^0\rangle=\operatorname{diag}\left\{\ii \asf^0_1,-\ii \asf^0_1, \ldots, \ii \asf^0_n,-\ii \asf^0_n\right\}
\end{equation}
and
\begin{equation}
\langle\phi^1\rangle=\operatorname{diag}\left\{\ii \asf^1_1,-\ii \asf^1_1, \ldots, \ii \asf^1_n,-\ii \asf^1_n\right\}.
\end{equation}
Rename the instanton counting parameter $\mathfrak{q}$ the desired Seiberg--Witten curve is
\begin{equation}
\label{eq:sw_D}
    y+\frac{\mathfrak{q}}{y}=\frac{\prod_{\alpha=1}^{n_0}(x^2-(\asf^0_\alpha)^2)}{\prod_{\alpha=1}^{n_1}(x^2-(\asf^1_\alpha)^2)+2/x^2}.
\end{equation}

For theories with $G=\mathrm{SpO}(n_0|2n_1)$, the corresponding Seiberg--Witten curve can be obtained similiarly as
\begin{equation}
\label{eq:sw_D_alt}
    y+\frac{\mathfrak{q}}{y}=\frac{\prod_{\alpha=1}^{n_0}(x^2-(\asf^1_\alpha)^2)+2/x^2}{\prod_{\alpha=1}^{n_1}(x^2-(\asf^0_\alpha)^2)}.
\end{equation}

This result is consistent with the result from brane construction by considering the gauging process from an orthosymplectic supergroup theory into an ordinary SQCD, see \eqref{SW_brane}.

\subsection{Seiberg--Witten Curve for $B_{n_0|n_1}$}
\label{sec:sw_Bsg}
\begin{figure}[htbp]
	\centering
	\includegraphics{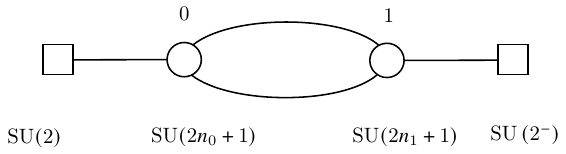}
	\caption{The embedding of $B_{n_0|n_1}$ supergroup gauge theory into a unitary quiver.}
	\label{fig:a1b}
\end{figure}
The $B_{n_0|n_1}$ case is a little bit different. To satisfy the supergroup condition, we need to change the second quiver node to $\SU(2n_1+1)$, as shown in fig. \ref{fig:a1b}. In this case the gauge polynomials are
\begin{equation}
\begin{aligned}
    \mathsf{T}_0 &= \prod_{\alpha=1}^{n_0}(x^2-(\asf^0_\alpha)^2)\\
    \mathsf{T}_1 &= x\prod_{\alpha=1}^{n_1}(x^2-(\asf^1_\alpha)^2)+\frac{2}{x}.
\end{aligned}
\end{equation}
with $P_0(x)=x$,
\begin{equation}
\langle\phi^0\rangle=\operatorname{diag}\left\{\ii \asf^0_1,-\ii \asf^0_1, \ldots, \ii \asf^0_n,-\ii \asf^0_n\right\}
\end{equation}
and
\begin{equation}
\langle\phi^1\rangle=\operatorname{diag}\left\{\ii \asf^1_1,-\ii \asf^1_1, \ldots, \ii \asf^1_n,-\ii \asf^1_n,0\right\}.
\end{equation}
The desired Seiberg--Witten curve is then
\begin{equation}
    y+\frac{\mathfrak{q}}{y}=\frac{\prod_{\alpha=1}^{n_0}(x^2-(\asf^0_\alpha)^2)}{x\prod_{\alpha=1}^{n_1}(x^2-(\asf^1_\alpha)^2)+2/x}.
\end{equation}
And for the $\mathrm{SpO}(n_0|2n_1+1)$ case, we have then
\begin{equation}
    y+\frac{\mathfrak{q}}{y}=\frac{x\prod_{\alpha=1}^{n_0}(x^2-(\asf^1_\alpha)^2)+2/x}{\prod_{\alpha=1}^{n_1}(x^2-(\asf^0_\alpha)^2)}.
\end{equation}

\section{Brane Construction}
\label{sec:brane_con}
\subsection{Positive and Negative Branes}
\label{sec:PNB}
Supergroup gauge theories can be constructed with the so-called \textit{negative branes} or \textit{ghost branes} \cite{Vafa:2001qf,Okuda:2006fb,Dijkgraaf:2016lym}. Using the Hanany-Witten construction, an 4d $\mathcal{N}=2$ super Yang--Mills theory with $\mathrm{U}(n_0|n_1)$ gauge group can be realized as a worldvolume theory of positive and negative branes, denoted $\mathrm{D4}^+$ and $\mathrm{D4}^-$, suspended between two separated $\mathrm{NS5}$ branes in the context of type IIA superstring theory. 

The following diagram shows the explicit construction. The horizontal solid and dotted lines represent the $\mathrm{D4}^+$ and $\mathrm{D4}^-$ branes:
\begin{align}
\tikzset{every picture/.style={line width=0.75pt}} 
\begin{tikzpicture}[x=0.75pt,y=0.75pt,yscale=-1,xscale=1,baseline=(current bounding box.center)]

\draw    (165.39,169.8) -- (165.2,199.2) ;
\draw [shift={(165.4,167.8)}, rotate = 90.36] [color={rgb, 255:red, 0; green, 0; blue, 0 }  ][line width=0.75]    (10.93,-3.29) .. controls (6.95,-1.4) and (3.31,-0.3) .. (0,0) .. controls (3.31,0.3) and (6.95,1.4) .. (10.93,3.29)   ;
\draw    (165.2,199.2) -- (192.8,199.2) ;
\draw [shift={(194.8,199.2)}, rotate = 180] [color={rgb, 255:red, 0; green, 0; blue, 0 }  ][line width=0.75]    (10.93,-3.29) .. controls (6.95,-1.4) and (3.31,-0.3) .. (0,0) .. controls (3.31,0.3) and (6.95,1.4) .. (10.93,3.29)   ;
\draw    (165.2,199.2) -- (189.81,173.84) ;
\draw [shift={(191.2,172.4)}, rotate = 134.13] [color={rgb, 255:red, 0; green, 0; blue, 0 }  ][line width=0.75]    (10.93,-3.29) .. controls (6.95,-1.4) and (3.31,-0.3) .. (0,0) .. controls (3.31,0.3) and (6.95,1.4) .. (10.93,3.29)   ;
\draw    (258.54,107.73) -- (258.6,217.8) ;
\draw    (368.21,129.87) -- (258.14,129.93) ;
\draw    (368.61,107.67) -- (368.67,217.74) ;
\draw    (368.21,155.47) -- (258.14,155.53) ;
\draw  [dash pattern={on 0.84pt off 2.51pt}]  (368.61,178.07) -- (258.54,178.08) ;
\draw  [dash pattern={on 0.84pt off 2.51pt}]  (369.01,200.87) -- (258.94,200.88) ;

\draw (156.2,151) node [anchor=north west][inner sep=0.75pt]    {$45$};
\draw (196.2,190.6) node [anchor=north west][inner sep=0.75pt]    {$6$};
\draw (185,155.8) node [anchor=north west][inner sep=0.75pt]    {$789$};
\draw (244.13,88.87) node [anchor=north west][inner sep=0.75pt]   [align=left] {NS5};
\draw (382.73,127.03) node [anchor=north west][inner sep=0.75pt]   [align=left] {$\displaystyle n_{0}$ D4$\displaystyle ^{+}$};
\draw (382.9,182.83) node [anchor=north west][inner sep=0.75pt]   [align=left] {$\displaystyle n_{1}$ D4$\displaystyle ^{-}$};
\draw (315.73,136.77) node [anchor=north west][inner sep=0.75pt]  [rotate=-90] [align=left] {...};
\draw (315.73,183.03) node [anchor=north west][inner sep=0.75pt]  [rotate=-90] [align=left] {...};
\draw (354.53,88.07) node [anchor=north west][inner sep=0.75pt]   [align=left] {NS5};

\end{tikzpicture}
\end{align}
From \cite{Dijkgraaf:2016lym} we know that we can annihilate the negative branes through gauging as follows:
\begin{align}
\label{eq:gauging}
\tikzset{every picture/.style={line width=0.75pt}} 
\begin{tikzpicture}[x=0.75pt,y=0.75pt,yscale=-1,xscale=1,baseline=(current bounding box.center)]

\draw    (115.15,93.42) -- (115.18,159.95) ;
\draw    (181.44,106.8) -- (114.91,106.86) ;
\draw    (181.68,93.38) -- (181.72,159.92) ;
\draw    (181.44,116.84) -- (114.91,116.9) ;
\draw  [dash pattern={on 0.84pt off 2.51pt}]  (181.7,126.68) -- (115.17,126.69) ;
\draw  [dash pattern={on 0.84pt off 2.51pt}]  (181.92,134.91) -- (115.39,134.91) ;
\draw    (200.36,144.28) -- (98.5,144.34) ;
\draw    (200.36,154.31) -- (98.5,154.37) ;
\draw    (246.74,93.62) -- (246.78,160.15) ;
\draw    (313.03,107) -- (246.5,107.06) ;
\draw    (313.28,93.58) -- (313.31,160.11) ;
\draw    (313.03,117.03) -- (246.5,117.09) ;
\draw  [dash pattern={on 0.84pt off 2.51pt}]  (313.29,126.88) -- (246.76,126.88) ;
\draw  [dash pattern={on 0.84pt off 2.51pt}]  (313.52,135.11) -- (246.98,135.11) ;
\draw    (246.42,144.48) -- (230.09,144.54) ;
\draw    (246.42,154.51) -- (230.09,154.57) ;
\draw    (329.53,143.87) -- (313.21,143.93) ;
\draw    (329.53,153.91) -- (313.21,153.97) ;
\draw    (313.21,139.34) -- (247.02,139.4) ;
\draw    (313.21,149.37) -- (247.02,149.43) ;
\draw    (375.71,93.21) -- (375.75,159.74) ;
\draw    (442,106.59) -- (375.47,106.65) ;
\draw    (442.24,93.17) -- (442.28,159.71) ;
\draw    (442,116.63) -- (375.47,116.69) ;
\draw    (375.38,144.07) -- (359.06,144.13) ;
\draw    (375.38,154.11) -- (359.06,154.17) ;
\draw    (458.5,143.47) -- (442.18,143.53) ;
\draw    (458.5,153.5) -- (442.18,153.56) ;
\draw    (201.16,119.93) .. controls (202.83,118.27) and (204.49,118.27) .. (206.16,119.94) .. controls (207.83,121.61) and (209.49,121.61) .. (211.16,119.95) .. controls (212.83,118.29) and (214.5,118.3) .. (216.16,119.97) -- (220.5,119.98) -- (228.5,120) ;
\draw [shift={(230.5,120)}, rotate = 180.13] [fill={rgb, 255:red, 0; green, 0; blue, 0 }  ][line width=0.08]  [draw opacity=0] (12,-3) -- (0,0) -- (12,3) -- cycle    ;
\draw    (330.94,119.83) .. controls (332.65,118.2) and (334.31,118.23) .. (335.94,119.94) .. controls (337.57,121.65) and (339.23,121.69) .. (340.94,120.06) .. controls (342.65,118.43) and (344.31,118.47) .. (345.93,120.18) -- (349.5,120.26) -- (357.5,120.45) ;
\draw [shift={(359.5,120.5)}, rotate = 181.35] [fill={rgb, 255:red, 0; green, 0; blue, 0 }  ][line width=0.08]  [draw opacity=0] (12,-3) -- (0,0) -- (12,3) -- cycle    ;

\end{tikzpicture}
\end{align}
If we gauge this supergroup gauge theory\footnote{This dose not imply the equivalence of these two kind of theories!}, we get a $\mathrm{U}(n_0)$ gauge theory with $n^{\mathrm{f}} = 2n_1$ flavors. This serves as a method to obtain the Seiberg--Witten curve for supergroup geuge theory because this gauging process does not change Coulomb branch geometry. We remark that such a reduction is possible only at the special locus of the Coulomb branch of the moduli space of vacua, and this does not mean the agreement of the moduli spaces themselves.

Similiar idea is also used by \cite{Dijkgraaf:2016lym} to explain results in \cite{Mikhaylov:2014aoa}, in which $\mathrm{U}(n_0|n_1)$ Chern-Simons theory on the three dimensional boundary is given by $n_0$ and $n_1$ D3-branes ending on opposite sides of an NS5-brane. We can first consider $n_0$ D3-branes and $n_1$ negative D3-branes on the same side. This leads to an $\mathcal{N} = 4$ $\mathrm{U}(n_0|n_1)$ gauge theory which gives a $\mathrm{U}(n_0|n_1)$ Chern-Simons theory on the boundary. Adding $n_1$ D3-branes passing through the NS5-branes and coinciding with the negative D3-branes will reover the boundary $\mathrm{U}(n_0|n_1)$ Chern-Simons theory, as shown schematically below:
\begin{align}
\tikzset{every picture/.style={line width=0.75pt}} 
\begin{tikzpicture}[x=0.75pt,y=0.75pt,yscale=-1,xscale=1,baseline=(current bounding box.center)]

\draw    (140,109.5) -- (140.18,179.95) ;
\draw    (189.64,124.8) -- (139.82,124.86) ;
\draw    (189.64,134.84) -- (139.82,134.9) ;
\draw  [dash pattern={on 0.84pt off 2.51pt}]  (189.83,144.68) -- (140.02,144.69) ;
\draw  [dash pattern={on 0.84pt off 2.51pt}]  (190,152.91) -- (140.18,152.91) ;
\draw    (189.5,162.28) -- (90.5,162.34) ;
\draw    (189.5,172.31) -- (90.5,172.37) ;
\draw    (201.66,140.43) .. controls (203.33,138.77) and (204.99,138.77) .. (206.66,140.44) .. controls (208.33,142.11) and (209.99,142.11) .. (211.66,140.45) .. controls (213.33,138.79) and (215,138.8) .. (216.66,140.47) -- (221,140.48) -- (229,140.5) ;
\draw [shift={(231,140.5)}, rotate = 180.13] [fill={rgb, 255:red, 0; green, 0; blue, 0 }  ][line width=0.08]  [draw opacity=0] (12,-3) -- (0,0) -- (12,3) -- cycle    ;
\draw    (351.44,140.33) .. controls (353.15,138.7) and (354.81,138.73) .. (356.44,140.44) .. controls (358.07,142.15) and (359.73,142.19) .. (361.44,140.56) .. controls (363.15,138.93) and (364.81,138.97) .. (366.43,140.68) -- (370,140.76) -- (378,140.95) ;
\draw [shift={(380,141)}, rotate = 181.35] [fill={rgb, 255:red, 0; green, 0; blue, 0 }  ][line width=0.08]  [draw opacity=0] (12,-3) -- (0,0) -- (12,3) -- cycle    ;
\draw    (289.5,110.5) -- (289.68,180.95) ;
\draw    (339.14,125.8) -- (289.32,125.86) ;
\draw    (339.14,135.84) -- (289.32,135.9) ;
\draw  [dash pattern={on 0.84pt off 2.51pt}]  (339.33,145.68) -- (289.52,145.69) ;
\draw  [dash pattern={on 0.84pt off 2.51pt}]  (339.5,153.91) -- (289.68,153.91) ;
\draw    (289.5,163.28) -- (241,163.34) ;
\draw    (289.5,173.31) -- (241,173.37) ;
\draw    (339.5,158.78) -- (289.5,158.84) ;
\draw    (339.5,168.81) -- (289.5,168.87) ;
\draw    (439.5,109.5) -- (439.68,179.95) ;
\draw    (489.64,135.3) -- (439.82,135.36) ;
\draw    (489.64,145.34) -- (439.82,145.4) ;
\draw    (440,152.28) -- (391.5,152.34) ;
\draw    (440,162.31) -- (391.5,162.37) ;

\draw (124.63,189.87) node [anchor=north west][inner sep=0.75pt]   [align=left] {NS5};
\draw (106.73,118.53) node [anchor=north west][inner sep=0.75pt]   [align=left] {D3$\displaystyle ^{+}$};
\draw (106.9,143.33) node [anchor=north west][inner sep=0.75pt]   [align=left] {D3$\displaystyle ^{-}$};

\end{tikzpicture}
\end{align}
We will show in section \ref{sec:ospbrane} that this kind of phenomena is also consistent with the Hanany-Witten construction of orthosymplectic supergroup gauge theory.

\subsection{Brane Constructions for Orthosymplectic Supergroups}
\label{sec:ospbrane}
\subsubsection{Pure Gauge Theory}
In analogy with the method adopted in \cite{Dijkgraaf:2016lym}, we put $n_0$ D4-branes and $n_1$ negative D4-branes between the NS5-branes and introduce an O4-plane\footnote{For a brief review on O4-planes, see appendix \ref{app:op}.}:
\begin{align}
\label{eq:brane_OSp}
\tikzset{every picture/.style={line width=0.75pt}} 
\begin{tikzpicture}[x=0.75pt,y=0.75pt,yscale=-1,xscale=1,baseline=(current bounding box.center)]

\draw    (250.54,109.73) -- (250.6,219.8) ;
\draw [color={rgb, 255:red, 47; green, 157; blue, 175 }  ,draw opacity=1 ] [dash pattern={on 4.5pt off 4.5pt}]  (250.6,219.8) -- (360.67,219.74) ;
\draw    (360.21,131.87) -- (250.14,131.93) ;
\draw    (360.61,109.67) -- (360.67,219.74) ;
\draw    (360.21,157.47) -- (250.14,157.53) ;
\draw  [dash pattern={on 0.84pt off 2.51pt}]  (360.61,180.07) -- (250.54,180.08) ;
\draw  [dash pattern={on 0.84pt off 2.51pt}]  (361.01,202.87) -- (250.94,202.88) ;
\draw [color={rgb, 255:red, 177; green, 0; blue, 0 }  ,draw opacity=1 ] [dash pattern={on 4.5pt off 4.5pt}]  (221.4,220.2) -- (250.6,219.8) ;
\draw [color={rgb, 255:red, 177; green, 0; blue, 0 }  ,draw opacity=1 ] [dash pattern={on 4.5pt off 4.5pt}]  (360.67,219.74) -- (389.87,219.34) ;

\draw (236.13,90.87) node [anchor=north west][inner sep=0.75pt]   [align=left] {NS5};
\draw (290.8,226.4) node [anchor=north west][inner sep=0.75pt]   [align=left] {O4$\displaystyle ^{-}$};
\draw (374.73,129.03) node [anchor=north west][inner sep=0.75pt]   [align=left] {$\displaystyle n_{0}$ D4$\displaystyle ^{+}$};
\draw (374.9,184.83) node [anchor=north west][inner sep=0.75pt]   [align=left] {$\displaystyle n_{1}$ D4$\displaystyle ^{-}$};
\draw (307.73,138.77) node [anchor=north west][inner sep=0.75pt]  [rotate=-90] [align=left] {...};
\draw (307.73,185.03) node [anchor=north west][inner sep=0.75pt]  [rotate=-90] [align=left] {...};
\draw (346.53,90.07) node [anchor=north west][inner sep=0.75pt]   [align=left] {NS5};
\draw (212.4,226.4) node [anchor=north west][inner sep=0.75pt]   [align=left] {O4$\displaystyle ^{+}$};
\draw (369.6,226) node [anchor=north west][inner sep=0.75pt]   [align=left] {O4$\displaystyle ^{+}$};

\end{tikzpicture}
\end{align}
We also note that the O4$^\pm$ plane will change its parity to O4$^\mp$ plane when crossing an NS5 brane. This gives rise to a $D_{n_0|n_1}$ or $\OSp(2n_0|n_1)$ supergroup gauge theory. For $B_{n_0|n_1}$ supergroup gauge theory, we need a half D4$^+$-brane coinciding with the O4$^-$-plane also. The SpO gauge theories can be constructed accordingly by replacing O4$^\pm$-plane by O4$^\mp$-plane.

We can check the consistency of this construction from several aspects. The easiest is to set $n_0$ or $n_1$ to be zero, and compare the bosonic subgroups. If $n_1=0$, this brane system gives rise to an $\SO(n_0)$ gauge theory. The effect of O4$^\pm$-planes acts on negative D4$^-$-branes is the same as O4$^\mp$-planes acts on ordiary D4$^+$-branes due to the $\Omega$ action \cite[sec.~2.3]{Okuda:2006fb}. Thus if $n_0=0$, it corresponds to an $\Sp(n_1)$ gauge theory, which resembles the $\SO(n_0)\times\Sp(n_1)$ bosonic subgroup.

\subsubsection{Adding Flavors}
Ordinary and negative flavor branes can be added to the brane system, as shown in the following:
\begin{align}
\tikzset{every picture/.style={line width=0.75pt}} 
\begin{tikzpicture}[x=0.75pt,y=0.75pt,yscale=-1,xscale=1,baseline=(current bounding box.center)]

\draw    (228.91,104.46) -- (228.96,190.26) ;
\draw [color={rgb, 255:red, 47; green, 157; blue, 175 }  ,draw opacity=1 ] [dash pattern={on 4.5pt off 4.5pt}]  (228.96,190.26) -- (314.77,190.22) ;
\draw    (314.41,121.72) -- (228.6,121.78) ;
\draw    (314.72,104.41) -- (314.77,190.22) ;
\draw    (314.41,141.68) -- (228.6,141.74) ;
\draw  [dash pattern={on 0.84pt off 2.51pt}]  (314.72,159.29) -- (228.91,159.3) ;
\draw  [dash pattern={on 0.84pt off 2.51pt}]  (315.03,177.07) -- (229.22,177.07) ;
\draw [color={rgb, 255:red, 177; green, 0; blue, 0 }  ,draw opacity=1 ] [dash pattern={on 4.5pt off 4.5pt}]  (206.2,190.58) -- (228.96,190.26) ;
\draw [color={rgb, 255:red, 177; green, 0; blue, 0 }  ,draw opacity=1 ] [dash pattern={on 4.5pt off 4.5pt}]  (314.77,190.22) -- (337.53,189.91) ;
\draw    (338.63,128.05) -- (314.36,128.11) ;
\draw    (339.41,148.47) -- (315.14,148.53) ;
\draw  [dash pattern={on 0.84pt off 2.51pt}]  (338.33,167.76) -- (314.05,167.76) ;
\draw    (228.71,128.44) -- (204.43,128.5) ;
\draw    (229.1,149.25) -- (204.82,149.31) ;
\draw  [dash pattern={on 0.84pt off 2.51pt}]  (228.4,168.15) -- (204.13,168.15) ;
\draw   (448.5,147.2) .. controls (448.5,141.13) and (453.43,136.2) .. (459.5,136.2) .. controls (465.58,136.2) and (470.51,141.12) .. (470.51,147.2) .. controls (470.51,153.27) and (465.59,158.2) .. (459.51,158.2) .. controls (453.44,158.21) and (448.51,153.28) .. (448.5,147.2) -- cycle ;
\draw  [fill={rgb, 255:red, 0; green, 0; blue, 0 }  ,fill opacity=1 ] (550.25,156.32) -- (530.81,156.32) -- (530.81,136.88) -- (550.25,136.88) -- cycle ;
\draw    (530.43,147.07) -- (470.51,147.2) ;

\draw (214.38,87.99) node [anchor=north west][inner sep=0.75pt]   [align=left] {NS5};
\draw (256.77,194.32) node [anchor=north west][inner sep=0.75pt]   [align=left] {O4$\displaystyle ^{-}$};
\draw (300.44,87.36) node [anchor=north west][inner sep=0.75pt]   [align=left] {NS5};
\draw (195.65,193.54) node [anchor=north west][inner sep=0.75pt]   [align=left] {O4$\displaystyle ^{+}$};
\draw (318.2,193.22) node [anchor=north west][inner sep=0.75pt]   [align=left] {O4$\displaystyle ^{+}$};
\draw (339.75,127.76) node [anchor=north west][inner sep=0.75pt]   [align=left] {D4$\displaystyle ^{+}$};
\draw (339.75,160.51) node [anchor=north west][inner sep=0.75pt]   [align=left] {D4$\displaystyle ^{-}$};
\draw (417,172.9) node [anchor=north west][inner sep=0.75pt]    {$\mathrm{OSp}( n_{0} |n_{1})$};
\draw (510.5,169) node [anchor=north west][inner sep=0.75pt]   [align=left] {SpO$\displaystyle \left( n_{0}^{\mathrm{f}} |n_{1}^{\mathrm{f}}\right)$};
\draw (391,140.9) node [anchor=north west][inner sep=0.75pt]    {$\leadsto $};

\end{tikzpicture}
\end{align}
Here the parity of the O4$^\pm$-plane changes parity when crossing the NS5-brane. Thus we can conclude that the flavor group of an OSp gauge theory preserves an SpO type flavor symmetry. This is also consistent with sec. \ref{sec:hyper} from global symmetry.

\subsubsection{Gauging Trick}
We can also perform gauging for this brane system, which schematically means applying similar operation in \eqref{eq:gauging} with O4-plane:
\begin{align}
\tikzset{every picture/.style={line width=0.75pt}} 
\begin{tikzpicture}[x=0.75pt,y=0.75pt,yscale=-1,xscale=1,baseline=(current bounding box.center)]

\draw    (135.15,97.27) -- (135.18,163.8) ;
\draw    (201.44,110.65) -- (134.91,110.71) ;
\draw    (201.68,97.23) -- (201.72,163.77) ;
\draw    (201.44,120.69) -- (134.91,120.75) ;
\draw  [dash pattern={on 0.84pt off 2.51pt}]  (201.7,130.53) -- (135.17,130.54) ;
\draw  [dash pattern={on 0.84pt off 2.51pt}]  (201.92,138.76) -- (135.39,138.76) ;
\draw    (220.36,148.13) -- (118.5,148.19) ;
\draw    (220.36,158.16) -- (118.5,158.22) ;
\draw    (266.74,97.47) -- (266.78,164) ;
\draw    (333.03,110.85) -- (266.5,110.91) ;
\draw    (333.28,97.43) -- (333.31,163.96) ;
\draw    (333.03,120.88) -- (266.5,120.94) ;
\draw  [dash pattern={on 0.84pt off 2.51pt}]  (333.29,130.73) -- (266.76,130.73) ;
\draw  [dash pattern={on 0.84pt off 2.51pt}]  (333.52,138.96) -- (266.98,138.96) ;
\draw    (266.42,148.33) -- (250.09,148.39) ;
\draw    (266.42,158.36) -- (250.09,158.42) ;
\draw    (349.53,147.72) -- (333.21,147.78) ;
\draw    (349.53,157.76) -- (333.21,157.82) ;
\draw    (333.21,143.19) -- (267.02,143.25) ;
\draw    (333.21,153.22) -- (267.02,153.28) ;
\draw    (395.71,97.06) -- (395.75,163.59) ;
\draw    (462,110.44) -- (395.47,110.5) ;
\draw    (462.24,97.02) -- (462.28,163.56) ;
\draw    (462,120.48) -- (395.47,120.54) ;
\draw    (395.38,147.92) -- (379.06,147.98) ;
\draw    (395.38,157.96) -- (379.06,158.02) ;
\draw    (478.5,147.32) -- (462.18,147.38) ;
\draw    (478.5,157.35) -- (462.18,157.41) ;
\draw    (221.16,123.78) .. controls (222.83,122.12) and (224.49,122.12) .. (226.16,123.79) .. controls (227.83,125.46) and (229.49,125.46) .. (231.16,123.8) .. controls (232.83,122.14) and (234.5,122.15) .. (236.16,123.82) -- (240.5,123.83) -- (248.5,123.85) ;
\draw [shift={(250.5,123.85)}, rotate = 180.13] [fill={rgb, 255:red, 0; green, 0; blue, 0 }  ][line width=0.08]  [draw opacity=0] (12,-3) -- (0,0) -- (12,3) -- cycle    ;
\draw    (350.94,123.68) .. controls (352.65,122.05) and (354.31,122.08) .. (355.94,123.79) .. controls (357.57,125.5) and (359.23,125.54) .. (360.94,123.91) .. controls (362.65,122.28) and (364.31,122.32) .. (365.93,124.03) -- (369.5,124.11) -- (377.5,124.3) ;
\draw [shift={(379.5,124.35)}, rotate = 181.35] [fill={rgb, 255:red, 0; green, 0; blue, 0 }  ][line width=0.08]  [draw opacity=0] (12,-3) -- (0,0) -- (12,3) -- cycle    ;
\draw [color={rgb, 255:red, 177; green, 0; blue, 0 }  ,draw opacity=1 ] [dash pattern={on 4.5pt off 4.5pt}]  (135.18,163.8) -- (201.72,163.77) ;
\draw [color={rgb, 255:red, 47; green, 157; blue, 175 }  ,draw opacity=1 ] [dash pattern={on 4.5pt off 4.5pt}]  (201.72,163.77) -- (220.5,163.5) ;
\draw [color={rgb, 255:red, 47; green, 157; blue, 175 }  ,draw opacity=1 ] [dash pattern={on 4.5pt off 4.5pt}]  (119,164) -- (135.18,163.8) ;
\draw [color={rgb, 255:red, 177; green, 0; blue, 0 }  ,draw opacity=1 ] [dash pattern={on 4.5pt off 4.5pt}]  (266.78,164) -- (333.31,163.96) ;
\draw [color={rgb, 255:red, 177; green, 0; blue, 0 }  ,draw opacity=1 ] [dash pattern={on 4.5pt off 4.5pt}]  (395.75,163.59) -- (462.28,163.56) ;
\draw [color={rgb, 255:red, 47; green, 157; blue, 175 }  ,draw opacity=1 ] [dash pattern={on 4.5pt off 4.5pt}]  (333.31,163.96) -- (352.09,163.7) ;
\draw [color={rgb, 255:red, 47; green, 157; blue, 175 }  ,draw opacity=1 ] [dash pattern={on 4.5pt off 4.5pt}]  (462.28,163.56) -- (481.06,163.29) ;
\draw [color={rgb, 255:red, 47; green, 157; blue, 175 }  ,draw opacity=1 ] [dash pattern={on 4.5pt off 4.5pt}]  (250.59,164.2) -- (266.78,164) ;
\draw [color={rgb, 255:red, 47; green, 157; blue, 175 }  ,draw opacity=1 ] [dash pattern={on 4.5pt off 4.5pt}]  (379.56,163.79) -- (395.75,163.59) ;

\end{tikzpicture}
\end{align}
In our case, the blue dashed line corresponds to O4$^-$-plane. The gauged theory is an SQCD with orthogonal gauge group and symplectic flavor group. Once again, this gauging process is possible only at the special locus of the Coulomb branch and does not infer the agreement of the moduli spaces themselves.

In order to check the consistency of this construction, we can compare the Seiberg--Witten curve of the theory before and after gauging process. For sake of simplicity, consider an $\mathrm{SpO}(n_0|2n_1)$ supergroup gauge theory which gives an $\Sp(n_0)$ gauge theory with $n^{\mathrm{f}}=4n_1$ flavors. The OSp case can be treated similarly. We can check the consistency by taking the Seiberg--Witten curve \eqref{eq:sw_D} for $D_{n_0|n_1}$ and compare with the Seiberg--Witten curve for $\Sp(n_0)$ gauge theory with $\SO(4n_1)$ flavor group. For the latter we have (up to a change of variable):
\begin{equation}
y+\frac{\mathfrak{q}P(x)}{y}=\prod_{\alpha=1}^{n_0}\left(x^2-(\asf^1_{\alpha})^2\right)+\frac{2}{x^2},\quad P(x)=\prod_{\alpha=1}^{2n_1}\left(x^2-(\asf^0_{\alpha})^2\right) \,.
\end{equation}
then let $y\to y\sqrt{P(x)}$, we have
\begin{equation}
y+\frac{\mathfrak{q}}{y}=\frac{\prod_{\alpha=1}^{n_0}\left(x^2-(\asf^1_{\alpha})^2\right)+\frac{2}{x^2}}{\sqrt{P(x)}},\quad P(x)=\prod_{\alpha=1}^{2n_1}\left(x^2-(\asf^0_{\alpha})^2\right) \,.
\end{equation}
Since the masses of the fundamental hypermultiplets are pairwise equal to the Coulomb branch parameters, we can make the identification
\begin{equation}
\asf^1_{n_1+s}=\asf^1_s,\ s=1,...,n_1.
\end{equation}
this leads to
\begin{equation}
\label{SW_brane}
y+\frac{\mathfrak{q}}{y}=\frac{\prod_{\alpha=1}^{n_0}(x^2-(\asf^1_\alpha)^2)+2/x^2}{\prod_{\alpha=1}^{n_1}(x^2-(\asf^0_\alpha)^2)}
\end{equation}
This exactly gives \eqref{eq:sw_D_alt}. The same analysis also applies to the $B_{n_0|n_1}$ case.

Also, this will give an explanation to why the orthosymplectic Chern–Simons theory in \cite[sec.~5.2]{Mikhaylov:2014aoa} is two sided. Notice also when an O3-plane crosses an NS5-brane, its NS flux jumps. The gauge group jumps from orthogonal to symplectic accordingly. We can do the gauging as
\begin{align}
\tikzset{every picture/.style={line width=0.75pt}} 
\begin{tikzpicture}[x=0.75pt,y=0.75pt,yscale=-1,xscale=1,baseline=(current bounding box.center)]

\draw    (160,33.97) -- (160.18,104.42) ;
\draw    (209.64,49.27) -- (159.82,49.33) ;
\draw    (209.64,59.3) -- (159.82,59.36) ;
\draw  [dash pattern={on 0.84pt off 2.51pt}]  (209.83,69.15) -- (160.02,69.15) ;
\draw  [dash pattern={on 0.84pt off 2.51pt}]  (210,77.38) -- (160.18,77.38) ;
\draw    (209.5,86.75) -- (110.5,86.81) ;
\draw    (209.5,96.78) -- (110.5,96.84) ;
\draw    (221.66,64.9) .. controls (223.33,63.23) and (225,63.24) .. (226.66,64.91) .. controls (228.33,66.58) and (229.99,66.58) .. (231.66,64.92) .. controls (233.33,63.26) and (234.99,63.26) .. (236.66,64.93) -- (241,64.94) -- (249,64.96) ;
\draw [shift={(251,64.97)}, rotate = 180.13] [fill={rgb, 255:red, 0; green, 0; blue, 0 }  ][line width=0.08]  [draw opacity=0] (12,-3) -- (0,0) -- (12,3) -- cycle    ;
\draw    (371.44,64.79) .. controls (373.15,63.16) and (374.81,63.2) .. (376.44,64.91) .. controls (378.07,66.62) and (379.73,66.66) .. (381.44,65.03) .. controls (383.15,63.4) and (384.81,63.44) .. (386.43,65.15) -- (390,65.23) -- (398,65.42) ;
\draw [shift={(400,65.47)}, rotate = 181.35] [fill={rgb, 255:red, 0; green, 0; blue, 0 }  ][line width=0.08]  [draw opacity=0] (12,-3) -- (0,0) -- (12,3) -- cycle    ;
\draw    (309.5,34.97) -- (309.68,105.42) ;
\draw    (359.14,50.27) -- (309.32,50.33) ;
\draw    (359.14,60.3) -- (309.32,60.36) ;
\draw  [dash pattern={on 0.84pt off 2.51pt}]  (359.33,70.15) -- (309.52,70.15) ;
\draw  [dash pattern={on 0.84pt off 2.51pt}]  (359.5,78.38) -- (309.68,78.38) ;
\draw    (309.5,87.75) -- (261,87.81) ;
\draw    (309.5,97.78) -- (261,97.84) ;
\draw    (359.5,83.25) -- (309.5,83.31) ;
\draw    (359.5,93.28) -- (309.5,93.34) ;
\draw    (459.5,33.97) -- (459.68,104.42) ;
\draw    (509.64,59.77) -- (459.82,59.83) ;
\draw    (509.64,69.8) -- (459.82,69.86) ;
\draw    (460,76.75) -- (411.5,76.81) ;
\draw    (460,86.78) -- (411.5,86.84) ;
\draw [color={rgb, 255:red, 47; green, 157; blue, 175 }  ,draw opacity=1 ] [dash pattern={on 4.5pt off 4.5pt}]  (160.18,104.42) -- (111.68,104.42) ;
\draw [color={rgb, 255:red, 177; green, 0; blue, 0 }  ,draw opacity=1 ] [dash pattern={on 4.5pt off 4.5pt}]  (208.68,104.42) -- (160.18,104.42) ;
\draw [color={rgb, 255:red, 47; green, 157; blue, 175 }  ,draw opacity=1 ] [dash pattern={on 4.5pt off 4.5pt}]  (309.68,105.42) -- (261.18,105.42) ;
\draw [color={rgb, 255:red, 177; green, 0; blue, 0 }  ,draw opacity=1 ] [dash pattern={on 4.5pt off 4.5pt}]  (358.18,105.42) -- (309.68,105.42) ;
\draw [color={rgb, 255:red, 47; green, 157; blue, 175 }  ,draw opacity=1 ] [dash pattern={on 4.5pt off 4.5pt}]  (459.68,104.42) -- (411.18,104.42) ;
\draw [color={rgb, 255:red, 177; green, 0; blue, 0 }  ,draw opacity=1 ] [dash pattern={on 4.5pt off 4.5pt}]  (508.18,104.42) -- (459.68,104.42) ;

\draw (171.5,106.5) node [anchor=north west][inner sep=0.75pt]   [align=left] {O3$\displaystyle ^{+}$};
\draw (121.5,107.5) node [anchor=north west][inner sep=0.75pt]   [align=left] {O3$\displaystyle ^{-}$};

\end{tikzpicture}
\end{align}
and applying the same argument as we did around \eqref{eq:gauging} will give the desired result.

\subsubsection{Affine Quiver Realization}
In section \ref{sec:Supergroup_Gauge_Theory} we have claimed that supergroup gauge theories can be realized as affine quivers. From brane construction, we can also see the structure as follows:
\begin{align}
\tikzset{every picture/.style={line width=0.75pt}} 
\begin{tikzpicture}[x=0.75pt,y=0.75pt,yscale=-1,xscale=1,baseline=(current bounding box.center)]

\draw    (432.75,91.07) -- (432.81,187.64) ;
\draw [color={rgb, 255:red, 47; green, 157; blue, 175 }  ,draw opacity=1 ] [dash pattern={on 4.5pt off 4.5pt}]  (432.81,187.64) -- (529.38,187.59) ;
\draw    (528.98,110.49) -- (432.4,110.55) ;
\draw    (529.33,91.01) -- (529.38,187.59) ;
\draw    (528.98,132.95) -- (432.4,133.01) ;
\draw  [dash pattern={on 0.84pt off 2.51pt}]  (529.33,152.78) -- (432.75,152.79) ;
\draw  [dash pattern={on 0.84pt off 2.51pt}]  (529.68,172.79) -- (433.11,172.79) ;
\draw [color={rgb, 255:red, 177; green, 0; blue, 0 }  ,draw opacity=1 ] [dash pattern={on 4.5pt off 4.5pt}]  (407.19,187.99) -- (432.81,187.64) ;
\draw [color={rgb, 255:red, 177; green, 0; blue, 0 }  ,draw opacity=1 ] [dash pattern={on 4.5pt off 4.5pt}]  (529.38,187.59) -- (555,187.24) ;
\draw    (225.54,91.8) -- (225.59,188.37) ;
\draw [color={rgb, 255:red, 47; green, 157; blue, 175 }  ,draw opacity=1 ] [dash pattern={on 4.5pt off 4.5pt}]  (225.59,188.37) -- (322.17,188.32) ;
\draw    (321.76,111.22) -- (225.19,111.28) ;
\draw    (322.11,91.75) -- (322.17,188.32) ;
\draw    (321.76,133.69) -- (225.19,133.75) ;
\draw    (322.11,159.66) -- (225.54,159.66) ;
\draw    (322.47,179.66) -- (225.89,179.66) ;
\draw  [draw opacity=0] (225.89,179.66) .. controls (225.89,179.66) and (225.89,179.66) .. (225.89,179.66) .. controls (219.97,179.66) and (215.17,177.31) .. (215.17,174.4) .. controls (215.17,171.56) and (219.76,169.24) .. (225.5,169.14) -- (225.89,174.4) -- cycle ; \draw   (225.89,179.66) .. controls (225.89,179.66) and (225.89,179.66) .. (225.89,179.66) .. controls (219.97,179.66) and (215.17,177.31) .. (215.17,174.4) .. controls (215.17,171.56) and (219.76,169.24) .. (225.5,169.14) ;  
\draw  [draw opacity=0] (225.54,159.66) .. controls (225.54,159.66) and (225.54,159.66) .. (225.54,159.66) .. controls (219.62,159.66) and (214.82,157.3) .. (214.82,154.4) .. controls (214.82,151.49) and (219.62,149.13) .. (225.54,149.13) -- (225.54,154.4) -- cycle ; \draw   (225.54,159.66) .. controls (225.54,159.66) and (225.54,159.66) .. (225.54,159.66) .. controls (219.62,159.66) and (214.82,157.3) .. (214.82,154.4) .. controls (214.82,151.49) and (219.62,149.13) .. (225.54,149.13) ;  
\draw  [draw opacity=0] (322.11,149.13) .. controls (322.11,149.13) and (322.11,149.13) .. (322.11,149.13) .. controls (328.03,149.12) and (332.83,151.48) .. (332.83,154.38) .. controls (332.83,157.29) and (328.04,159.65) .. (322.11,159.66) -- (322.11,154.39) -- cycle ; \draw   (322.11,149.13) .. controls (322.11,149.13) and (322.11,149.13) .. (322.11,149.13) .. controls (328.03,149.12) and (332.83,151.48) .. (332.83,154.38) .. controls (332.83,157.29) and (328.04,159.65) .. (322.11,159.66) ;  
\draw  [draw opacity=0] (322.46,169.13) .. controls (322.46,169.13) and (322.46,169.13) .. (322.46,169.13) .. controls (328.38,169.13) and (333.18,171.48) .. (333.18,174.39) .. controls (333.19,177.3) and (328.39,179.66) .. (322.47,179.66) -- (322.46,174.4) -- cycle ; \draw   (322.46,169.13) .. controls (322.46,169.13) and (322.46,169.13) .. (322.46,169.13) .. controls (328.38,169.13) and (333.18,171.48) .. (333.18,174.39) .. controls (333.19,177.3) and (328.39,179.66) .. (322.47,179.66) ;  
\draw [color={rgb, 255:red, 245; green, 166; blue, 35 }  ,draw opacity=1 ]   (321.76,74.37) -- (225.19,74.38) ;
\draw [shift={(225.19,74.38)}, rotate = 360] [color={rgb, 255:red, 245; green, 166; blue, 35 }  ,draw opacity=1 ][line width=0.75]    (0,5.59) -- (0,-5.59)(10.93,-3.29) .. controls (6.95,-1.4) and (3.31,-0.3) .. (0,0) .. controls (3.31,0.3) and (6.95,1.4) .. (10.93,3.29)   ;
\draw [shift={(321.76,74.37)}, rotate = 180] [color={rgb, 255:red, 245; green, 166; blue, 35 }  ,draw opacity=1 ][line width=0.75]    (0,5.59) -- (0,-5.59)(10.93,-3.29) .. controls (6.95,-1.4) and (3.31,-0.3) .. (0,0) .. controls (3.31,0.3) and (6.95,1.4) .. (10.93,3.29)   ;
\draw [color={rgb, 255:red, 245; green, 166; blue, 35 }  ,draw opacity=1 ]   (322.11,218.15) -- (225.54,218.15) ;
\draw  [draw opacity=0] (225.54,218.15) .. controls (225.54,218.15) and (225.54,218.15) .. (225.54,218.15) .. controls (219.62,218.15) and (214.82,215.8) .. (214.82,212.89) .. controls (214.82,209.98) and (219.62,207.62) .. (225.54,207.62) -- (225.54,212.89) -- cycle ; \draw [color={rgb, 255:red, 245; green, 166; blue, 35 }  ,draw opacity=1 ]   (225.54,218.15) .. controls (219.62,218.15) and (214.82,215.8) .. (214.82,212.89) .. controls (214.82,209.98) and (219.62,207.62) .. (225.54,207.62) ; \draw [shift={(225.54,207.62)}, rotate = 169.5] [color={rgb, 255:red, 245; green, 166; blue, 35 }  ,draw opacity=1 ][line width=0.75]    (0,3.35) -- (0,-3.35)(6.56,-1.97) .. controls (4.17,-0.84) and (1.99,-0.18) .. (0,0) .. controls (1.99,0.18) and (4.17,0.84) .. (6.56,1.97)   ; 
\draw    (356.53,141.02) -- (398.99,141.02) ;
\draw [shift={(400.99,141.02)}, rotate = 180] [color={rgb, 255:red, 0; green, 0; blue, 0 }  ][line width=0.75]    (7.65,-2.3) .. controls (4.86,-0.97) and (2.31,-0.21) .. (0,0) .. controls (2.31,0.21) and (4.86,0.98) .. (7.65,2.3)   ;
\draw  [draw opacity=0][dash pattern={on 4.5pt off 4.5pt}] (322.17,188.32) .. controls (322.17,188.32) and (322.17,188.32) .. (322.17,188.32) .. controls (328.09,188.31) and (332.89,190.67) .. (332.89,193.58) .. controls (332.89,196.48) and (328.1,198.84) .. (322.17,198.85) -- (322.17,193.58) -- cycle ; \draw  [color={rgb, 255:red, 177; green, 0; blue, 0 }  ,draw opacity=1 ][dash pattern={on 4.5pt off 4.5pt}] (322.17,188.32) .. controls (322.17,188.32) and (322.17,188.32) .. (322.17,188.32) .. controls (328.09,188.31) and (332.89,190.67) .. (332.89,193.58) .. controls (332.89,196.48) and (328.1,198.84) .. (322.17,198.85) ;  
\draw  [draw opacity=0][dash pattern={on 4.5pt off 4.5pt}] (225.99,198.9) .. controls (225.99,198.9) and (225.99,198.9) .. (225.99,198.9) .. controls (220.07,198.9) and (215.27,196.54) .. (215.27,193.63) .. controls (215.27,190.79) and (219.86,188.47) .. (225.59,188.37) -- (225.99,193.63) -- cycle ; \draw  [color={rgb, 255:red, 177; green, 0; blue, 0 }  ,draw opacity=1 ][dash pattern={on 4.5pt off 4.5pt}] (225.99,198.9) .. controls (225.99,198.9) and (225.99,198.9) .. (225.99,198.9) .. controls (220.07,198.9) and (215.27,196.54) .. (215.27,193.63) .. controls (215.27,190.79) and (219.86,188.47) .. (225.59,188.37) ;  
\draw [color={rgb, 255:red, 177; green, 0; blue, 0 }  ,draw opacity=1 ] [dash pattern={on 4.5pt off 4.5pt}]  (225.99,198.9) -- (322.17,198.85) ;
\draw  [draw opacity=0] (322.21,207.62) .. controls (328.09,207.65) and (332.84,209.99) .. (332.84,212.89) .. controls (332.84,215.79) and (328.04,218.15) .. (322.11,218.15) -- (322.11,212.89) -- cycle ; \draw [color={rgb, 255:red, 245; green, 166; blue, 35 }  ,draw opacity=1 ]   (322.21,207.62) .. controls (328.09,207.65) and (332.84,209.99) .. (332.84,212.89) .. controls (332.84,215.79) and (328.04,218.15) .. (322.11,218.15) ;  \draw [shift={(322.21,207.62)}, rotate = 5.72] [color={rgb, 255:red, 245; green, 166; blue, 35 }  ,draw opacity=1 ][line width=0.75]    (0,3.35) -- (0,-3.35)(6.56,-1.97) .. controls (4.17,-0.84) and (1.99,-0.18) .. (0,0) .. controls (1.99,0.18) and (4.17,0.84) .. (6.56,1.97)   ;

\draw (264.35,79.84) node [anchor=north west][inner sep=0.75pt]  [color={rgb, 255:red, 245; green, 166; blue, 35 }  ,opacity=1 ]  {$L_{0}$};
\draw (262.94,224.08) node [anchor=north west][inner sep=0.75pt]  [color={rgb, 255:red, 245; green, 166; blue, 35 }  ,opacity=1 ]  {$L_{1}$};
\draw (346.9,117.57) node [anchor=north west][inner sep=0.75pt]    {$L_{1}\rightarrow -L_{0}$};

\end{tikzpicture}
\end{align}
First consider an $\widehat{A}_1$ cyclic quiver with $\OO(2n_0+\chi_n)\times\Sp(n_1)$ gauge nodes and perform the analytic continuation to the unphysical regime $g_0^2\to -g_1^2$, which in Hanany-Witten construction corresponds to setting $L_0\to-L_1$. Therefore the brane configuration after analytic continuation is just what we get in \eqref{eq:brane_OSp}.

From this affine quiver realization we can see its relation between an ordinary SQCD by considering its weak coupling limit as follows. Recall the supergroup condition in sec. \ref{sec:A1^_supergroup}:
\begin{equation}
   \tau_0+\tau_1=1
\end{equation}
or in the exponential form
\begin{equation}
    \mathfrak{q}_0\mathfrak{q}_1=1
\end{equation}
with $\mathfrak{q}_i=\exp(2\pi\ii \tau_i)$ and $\tau_i=\frac{\theta}{2\pi}+\frac{4\pi\ii}{g_i^2}$ the complexified Yang-Mills coupling. Set
\begin{equation}
    \mathfrak{Q}=\mathfrak{q}_0\mathfrak{q}_1=\exp(2\pi\ii \tau_{\mathrm{tot.}}).
\end{equation}
By a modular transform sending $\mathrm{Im}\,\tau_{\mathrm{tot.}}\to0$ we have then $\mathfrak{Q}=0$, which corresponds to the weak coupling limit. From brane perspective we can interpret it as:
\begin{align}
\input{figures/weak}
\end{align}
This leads to a ordinary SQCD. However, we should emphasize that this relation holds only at weak coupling limit and does not leads to an equivalence between ordinary gauge theories and supergroup gauge theories.

\subsubsection{One-to-Many Correspondence}
We notice that in the brane construction of supergroup gauge theories, the ordering of the positive and negative branes can have different choices. This is related to the ambiguity of the simple root decomposition of the supergroup. Thus we can classify different configurations by distinguished super Dynkin diagrams, see, e.g., \cite{Frappat:1996pb}. The rules is given as follows: We assign the bosonic (or ordinary) node to the neighboring pair of D4$^+$-D4$^+$ or D4$^-$-D4$^-$ branes, and the fermionic node is assigned to the neighboring pair of D4$^+$-D4$^-$ branes. This also works for external flavor branes. We note that from the topological string analysis \cite{Kimura:2020lmc}, the partition function does not depend on the ordering of positive and negative branes for unitary supergroup gauge theories. 

Here we present some examples of the correspondence rules. And for further explanation, we refer the readers to \cite[sec.~36]{Frappat:1996pb}. We also note that a \textit{canonical choice} of the super Dynkin diagram is to say it has a minimal number of fermonic nodes. In the following diagrams $\bigcirc$ stands for an bosonic node and {\scriptsize $\bigotimes$} the fermonic nodes while the red colour marks the canonical choice.

\paragraph{The $C_n$ Case}
\begin{align}
\tikzset{every picture/.style={line width=0.75pt}} 
\begin{tikzpicture}[x=0.75pt,y=0.75pt,yscale=-1,xscale=1,baseline=(current bounding box.center)]

\draw    (200.74,110.62) -- (200.78,177.15) ;
\draw    (267.03,127) -- (200.5,127.06) ;
\draw    (267.28,110.58) -- (267.31,177.11) ;
\draw  [dash pattern={on 0.84pt off 2.51pt}]  (266.29,144.88) -- (199.76,144.88) ;
\draw  [dash pattern={on 0.84pt off 2.51pt}]  (267.02,161.11) -- (200.48,161.11) ;
\draw [color={rgb, 255:red, 47; green, 157; blue, 175 }  ,draw opacity=1 ] [dash pattern={on 4.5pt off 4.5pt}]  (200.78,177.15) -- (267.31,177.11) ;
\draw    (340.74,110.62) -- (340.78,177.15) ;
\draw  [dash pattern={on 0.84pt off 2.51pt}]  (407.03,127) -- (340.5,127) ;
\draw    (407.28,110.58) -- (407.31,177.11) ;
\draw    (407.29,144.88) -- (340.76,144.88) ;
\draw  [dash pattern={on 0.84pt off 2.51pt}]  (407.02,161.11) -- (340.48,161.11) ;
\draw [color={rgb, 255:red, 47; green, 157; blue, 175 }  ,draw opacity=1 ] [dash pattern={on 4.5pt off 4.5pt}]  (340.78,177.15) -- (407.31,177.11) ;
\draw    (480.74,111.12) -- (480.78,177.65) ;
\draw    (547.28,111.08) -- (547.31,177.61) ;
\draw  [dash pattern={on 0.84pt off 2.51pt}]  (546.29,145.38) -- (479.76,145.38) ;
\draw    (547.02,161.61) -- (480.48,161.61) ;
\draw [color={rgb, 255:red, 47; green, 157; blue, 175 }  ,draw opacity=1 ] [dash pattern={on 4.5pt off 4.5pt}]  (480.78,177.65) -- (547.31,177.61) ;
\draw  [dash pattern={on 0.84pt off 2.51pt}]  (547.03,127) -- (480.5,127) ;
\draw  [color={rgb, 255:red, 177; green, 0; blue, 0 }  ,draw opacity=1 ] (177.43,134.75) .. controls (178.87,133.31) and (181.19,133.31) .. (182.62,134.75) .. controls (184.06,136.18) and (184.06,138.51) .. (182.62,139.94) .. controls (181.19,141.37) and (178.87,141.37) .. (177.43,139.94) .. controls (176,138.51) and (176,136.18) .. (177.43,134.75) -- cycle ;
\draw [color={rgb, 255:red, 177; green, 0; blue, 0 }  ,draw opacity=1 ]   (177.43,134.75) -- (182.62,139.94) ;
\draw [color={rgb, 255:red, 177; green, 0; blue, 0 }  ,draw opacity=1 ]   (182.62,134.75) -- (177.43,139.94) ;

\draw  [color={rgb, 255:red, 177; green, 0; blue, 0 }  ,draw opacity=1 ] (176.36,154.08) .. controls (176.36,152.05) and (178,150.41) .. (180.03,150.41) .. controls (182.06,150.41) and (183.7,152.05) .. (183.7,154.08) .. controls (183.7,156.11) and (182.06,157.75) .. (180.03,157.75) .. controls (178,157.75) and (176.36,156.11) .. (176.36,154.08) -- cycle ;
\draw [color={rgb, 255:red, 177; green, 0; blue, 0 }  ,draw opacity=1 ]   (178.5,167.05) -- (178.53,157.75)(181.5,167.06) -- (181.53,157.76) ;
\draw [shift={(180.02,158.8)}, rotate = 90.15] [color={rgb, 255:red, 177; green, 0; blue, 0 }  ,draw opacity=1 ][line width=0.75]    (6.56,-2.94) .. controls (4.17,-1.38) and (1.99,-0.4) .. (0,0) .. controls (1.99,0.4) and (4.17,1.38) .. (6.56,2.94)   ;
\draw  [color={rgb, 255:red, 177; green, 0; blue, 0 }  ,draw opacity=1 ] (176.33,170.73) .. controls (176.33,168.7) and (177.98,167.05) .. (180,167.05) .. controls (182.03,167.05) and (183.67,168.7) .. (183.67,170.73) .. controls (183.67,172.75) and (182.03,174.4) .. (180,174.4) .. controls (177.98,174.4) and (176.33,172.75) .. (176.33,170.73) -- cycle ;
\draw [color={rgb, 255:red, 177; green, 0; blue, 0 }  ,draw opacity=1 ]   (180.03,150.41) -- (180.15,141.02) ;
\draw [color={rgb, 255:red, 177; green, 0; blue, 0 }  ,draw opacity=1 ]   (458.35,165.92) -- (458.38,156.62)(461.35,165.93) -- (461.38,156.63) ;
\draw [shift={(459.87,157.67)}, rotate = 90.15] [color={rgb, 255:red, 177; green, 0; blue, 0 }  ,draw opacity=1 ][line width=0.75]    (6.56,-2.94) .. controls (4.17,-1.38) and (1.99,-0.4) .. (0,0) .. controls (1.99,0.4) and (4.17,1.38) .. (6.56,2.94)   ;
\draw [color={rgb, 255:red, 177; green, 0; blue, 0 }  ,draw opacity=1 ]   (459.88,149.28) -- (460,139.89) ;
\draw  [color={rgb, 255:red, 177; green, 0; blue, 0 }  ,draw opacity=1 ] (456.32,136.21) .. controls (456.32,134.19) and (457.97,132.54) .. (460,132.54) .. controls (462.02,132.54) and (463.67,134.19) .. (463.67,136.21) .. controls (463.67,138.24) and (462.02,139.89) .. (460,139.89) .. controls (457.97,139.89) and (456.32,138.24) .. (456.32,136.21) -- cycle ;
\draw  [color={rgb, 255:red, 177; green, 0; blue, 0 }  ,draw opacity=1 ] (457.39,150.49) .. controls (458.82,149.06) and (461.15,149.06) .. (462.58,150.49) .. controls (464.02,151.92) and (464.02,154.25) .. (462.58,155.68) .. controls (461.15,157.12) and (458.82,157.12) .. (457.39,155.68) .. controls (455.96,154.25) and (455.96,151.92) .. (457.39,150.49) -- cycle ;
\draw [color={rgb, 255:red, 177; green, 0; blue, 0 }  ,draw opacity=1 ]   (457.39,150.49) -- (462.58,155.68) ;
\draw [color={rgb, 255:red, 177; green, 0; blue, 0 }  ,draw opacity=1 ]   (462.58,150.49) -- (457.39,155.68) ;

\draw  [color={rgb, 255:red, 177; green, 0; blue, 0 }  ,draw opacity=1 ] (456.18,169.6) .. controls (456.18,167.57) and (457.83,165.93) .. (459.85,165.93) .. controls (461.88,165.93) and (463.52,167.57) .. (463.52,169.6) .. controls (463.52,171.62) and (461.88,173.27) .. (459.85,173.27) .. controls (457.83,173.27) and (456.18,171.62) .. (456.18,169.6) -- cycle ;
\draw  [color={rgb, 255:red, 0; green, 0; blue, 0 }  ,draw opacity=1 ] (318.53,133.63) .. controls (319.97,132.2) and (322.29,132.2) .. (323.73,133.63) .. controls (325.16,135.06) and (325.16,137.39) .. (323.73,138.82) .. controls (322.29,140.26) and (319.97,140.26) .. (318.53,138.82) .. controls (317.1,137.39) and (317.1,135.06) .. (318.53,133.63) -- cycle ;
\draw [color={rgb, 255:red, 0; green, 0; blue, 0 }  ,draw opacity=1 ]   (318.53,133.63) -- (323.73,138.82) ;
\draw [color={rgb, 255:red, 0; green, 0; blue, 0 }  ,draw opacity=1 ]   (323.73,133.63) -- (318.53,138.82) ;

\draw [color={rgb, 255:red, 0; green, 0; blue, 0 }  ,draw opacity=1 ]   (319.61,165.93) -- (319.63,156.63)(322.61,165.94) -- (322.63,156.64) ;
\draw [shift={(321.13,157.69)}, rotate = 90.15] [color={rgb, 255:red, 0; green, 0; blue, 0 }  ,draw opacity=1 ][line width=0.75]    (6.56,-2.94) .. controls (4.17,-1.38) and (1.99,-0.4) .. (0,0) .. controls (1.99,0.4) and (4.17,1.38) .. (6.56,2.94)   ;
\draw  [color={rgb, 255:red, 0; green, 0; blue, 0 }  ,draw opacity=1 ] (317.43,169.61) .. controls (317.43,167.58) and (319.08,165.94) .. (321.11,165.94) .. controls (323.13,165.94) and (324.78,167.58) .. (324.78,169.61) .. controls (324.78,171.64) and (323.13,173.28) .. (321.11,173.28) .. controls (319.08,173.28) and (317.43,171.64) .. (317.43,169.61) -- cycle ;
\draw [color={rgb, 255:red, 0; green, 0; blue, 0 }  ,draw opacity=1 ]   (321.13,149.29) -- (321.25,139.9) ;
\draw  [color={rgb, 255:red, 0; green, 0; blue, 0 }  ,draw opacity=1 ] (318.64,150.39) .. controls (320.08,148.96) and (322.4,148.96) .. (323.83,150.39) .. controls (325.27,151.83) and (325.27,154.15) .. (323.83,155.59) .. controls (322.4,157.02) and (320.08,157.02) .. (318.64,155.59) .. controls (317.21,154.15) and (317.21,151.83) .. (318.64,150.39) -- cycle ;
\draw [color={rgb, 255:red, 0; green, 0; blue, 0 }  ,draw opacity=1 ]   (318.64,150.39) -- (323.83,155.59) ;
\draw [color={rgb, 255:red, 0; green, 0; blue, 0 }  ,draw opacity=1 ]   (323.83,150.39) -- (318.64,155.59) ;

\end{tikzpicture}
\end{align}
This example shows $C_2=\OSp(2|2)$. The bottom node is always bosonic because the ``reflection'' effect of the orientifold plane. 

\paragraph{The $B_{n_0|n_1}$ Case}
\begin{align}
\input{figures/osp_2m1_2n}
\end{align}
This example shows $B_{1|2}=\OSp(3|2)$.

\paragraph{The $D_{n_0|n_1}$ Case}
\begin{align}
\input{figures/osp_2m_2n}
\end{align}
This example shows $D_{3|1}=\OSp(6|1)$. Note the numbering of the gauge nodes in the super Dynkin diagram.

\subsection{Quiver Constructions}
\label{sec:brane_quiver}
Similarly we can also construct quiver gauge theories by adding negative branes into the brane systems for O-Sp alternating quivers. Here we present the construction for $A_r$ or $D_r$ quivers. For the cases beyond (affine) $A_r$ or $D_r$ quivers, we may apply the approach shown in \cite{Kimura:2019gon} based on the formalism of \cite{Kimura:2015rgi,Kimura:2016dys,Kimura:2017hez}.

\subsubsection{$A_r$ quivers}
\begin{figure}[t]
	\centering
	\includegraphics{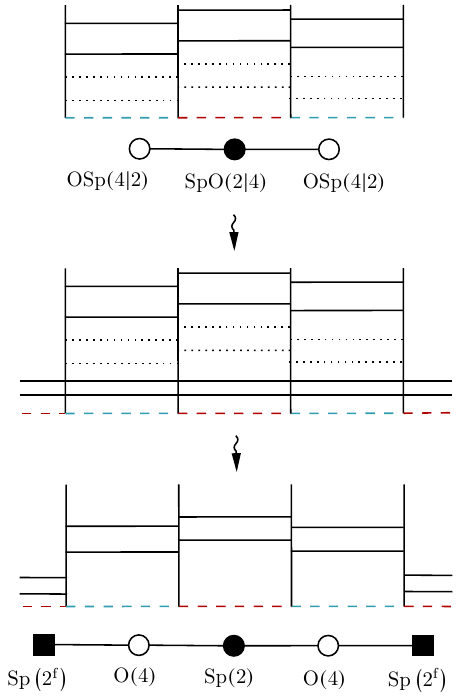}
	\caption{Reduction of $A_r$ quiver gauge theory via gauging.}
	\label{fig:a_quiver}
\end{figure}

The construction of linear or $A_r$ quivers is shown in fig. \ref{fig:a_quiver}. For sake of simplicity we assign the same gauge groups to all the gauge nodes. By gauging we can annihilate the negative branes and the resulting theory is the linear orthosymplectic quiver theory with alternating O/Sp gauge groups and two flavor groups at both ends. We can also consider the situation with different gauge groups assigned to each node. Then it is not possible to annihilate all the negative branes at the same time.

\subsubsection{$\widehat{A}_{r-1}$ quivers}
\begin{figure}[t]
	\centering
	\includegraphics{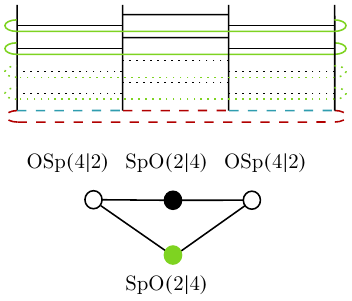}
	\caption{Brane configuration for affine $\widehat{A}_3$ quiver.}
	\label{fig:affine_quiver_a}
\end{figure}

We can also consider the cyclic quiver case $\widehat{A}_{r-1}$ as shown in fig. \ref{fig:affine_quiver_a}. We can align $r$ NS5 branes in periodic 6-direction and consider the D4$^\pm$ branes hanging in between. We remark that we cannot impose the fundamental hypermultiplet to this configuration using the external D4$^\pm$ branes.

\subsubsection{$D_r$ quivers}
\begin{figure}[t]
	\centering
	\includegraphics{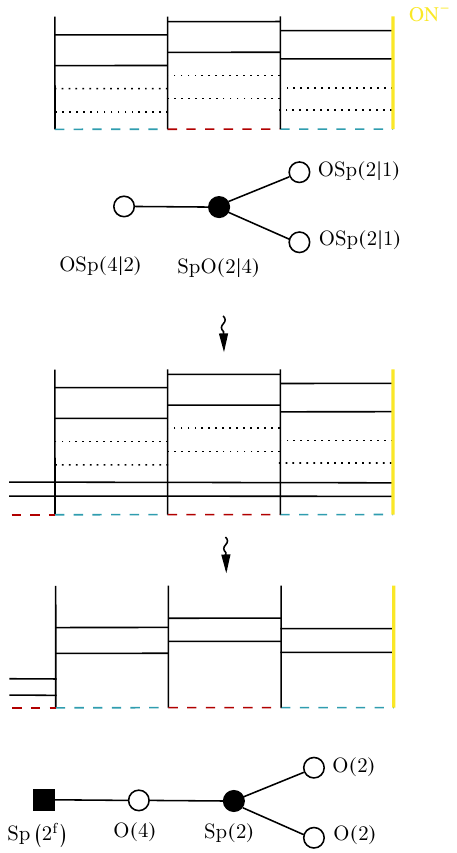}
	\caption{Reduction of $D_r$ quiver gauge theory via gauging.}
	\label{fig:d_quiver}
\end{figure}
We can also construct $D_r$ quivers according to \cite{Hanany:1999sj,Kapustin:1998fa} using $\mathrm{ON}^0$ plane, as shown in fig. \ref{fig:d_quiver}. It can be also reduce to $D_r$ quiver with non--supergroup gauge nodes. In addition, we can realize (affine) $\widehat{D}_r$ quivers by putting $\mathrm{ON}^0$ planes on the both ends of the brane system.

\section{Outlooks}
\label{sec:outlooks}
In this study, we presented the superinstanton counting for orthosymplectic supergroup gauge theories, derived the corresponding Seiberg--Witten geometry, and provided brane constructions for these theories. While we partially solved some of the problems raised in \cite[sec.~8.1]{Kimura:2019msw}, there is still much to be explored.

\paragraph{ADHM Constructions} 
Supersymmetric instantons can be constructed systematically using ADHM construction, and the instanton partition function is given by the equivariant localization on the instanton moduli space determined by the ADHM equations. However, we did not present the explicit mathematical construction of the superinstanton moduli space in this paper. A rigorous mathematical definition of the superinstanton moduli space (as well as for the unitary supergroup case) will be of both mathematical and physical interest.

\paragraph{Exceptional Supergroups}
The generalization of supergroup gauge theories with exceptional supergroups such as $G_3$ or $F_4$ may also have physical implications. 

\paragraph{Relations with Integrable Systems}
In \cite{Chen:2020rxu}, the relation of unitary supergroup with super integrable systems is explored. Similar relations are expected for the orthosymplectic case. The Lie superalgebra $D_{2|1;\alpha}$, which is a deformation of $D_{2|1}$, is of particular interest because the bosonic part of this algebra contains three copies of $\mathfrak{sl}(2)$, and the fermionic part transforms as a tri-fundamental representation. The origin of the parameter $\alpha$ is most likely coming from certain parameters from integrable systems.

\paragraph{Vertex formulism}
The lift-up to 5d $\mathcal{N}=1$ theory in the sense of \cite{Zafrir:2015ftn} has already been done for the unitary supergroup gauge theory in \cite{Kimura:2020lmc}. A supergroup generalization of the refined topological vertex \cite{Aganagic:2003db,Awata:2005fa,Iqbal:2007ii} is given by the anti-refined topological vertex (anti-vertex) introduced in \cite{Kimura:2020lmc}. The topological vertex formalism with O-planes is discussed in \cite{Kim:2017jqn,Bourgine:2017rik} and established in \cite{Hayashi:2020hhb,Nawata:2021dlk}, which is called the O-vertex. We can expect the existence of an \textit{anti-O-vertex} by combining the above ingredients.

We conjecture that the orthosymplectic supergroup gauge theory considered in this paper is dual to a $(p,q)$ 5-brane/negative brane web with an O-plane in type IIB string theory, via T-duality:
\begin{align}
\tikzset{every picture/.style={line width=0.75pt}} 
\begin{tikzpicture}[x=0.75pt,y=0.75pt,yscale=-1,xscale=1,baseline=(current bounding box.center)]

\draw    (201.74,112.62) -- (201.78,179.15) ;
\draw    (268.03,126) -- (201.5,126.06) ;
\draw    (268.28,112.58) -- (268.31,179.11) ;
\draw    (268.03,144.03) -- (201.5,144.09) ;
\draw  [dash pattern={on 0.84pt off 2.51pt}]  (267.29,155.88) -- (200.76,155.88) ;
\draw  [dash pattern={on 0.84pt off 2.51pt}]  (268.02,167.11) -- (201.48,167.11) ;
\draw    (332,149) -- (361,149)(332,152) -- (361,152) ;
\draw [shift={(369,150.5)}, rotate = 180] [color={rgb, 255:red, 0; green, 0; blue, 0 }  ][line width=0.75]    (10.93,-3.29) .. controls (6.95,-1.4) and (3.31,-0.3) .. (0,0) .. controls (3.31,0.3) and (6.95,1.4) .. (10.93,3.29)   ;
\draw [shift={(324,150.5)}, rotate = 0] [color={rgb, 255:red, 0; green, 0; blue, 0 }  ][line width=0.75]    (10.93,-3.29) .. controls (6.95,-1.4) and (3.31,-0.3) .. (0,0) .. controls (3.31,0.3) and (6.95,1.4) .. (10.93,3.29)   ;
\draw [color={rgb, 255:red, 47; green, 157; blue, 175 }  ,draw opacity=1 ] [dash pattern={on 4.5pt off 4.5pt}]  (201.78,179.15) -- (268.31,179.11) ;
\draw [color={rgb, 255:red, 47; green, 157; blue, 175 }  ,draw opacity=1 ] [dash pattern={on 4.5pt off 4.5pt}]  (418.67,180.17) -- (502.33,180.17) ;
\draw  [dash pattern={on 0.84pt off 2.51pt}]  (474.76,164.65) -- (502.33,180.17) ;
\draw  [dash pattern={on 0.84pt off 2.51pt}]  (446.24,164.65) -- (418.67,180.17) ;
\draw  [dash pattern={on 0.84pt off 2.51pt}]  (465.41,149.89) -- (474.76,164.65) ;
\draw  [dash pattern={on 0.84pt off 2.51pt}]  (454.66,150.27) -- (446.24,164.65) ;
\draw    (473.82,126.8) -- (465.41,142.45) ;
\draw    (446.71,126.43) -- (454.66,142.32) ;
\draw    (419.13,111.67) -- (446.71,126.43) ;
\draw    (500,112.42) -- (473.82,126.8) ;
\draw    (454.66,142.32) -- (454.66,150.27) ;
\draw    (465.72,142.32) -- (465.72,150.27) ;
\draw    (446.71,126.43) -- (473.82,126.8) ;
\draw    (454.66,142.32) -- (465.41,142.45) ;
\draw  [dash pattern={on 0.84pt off 2.51pt}]  (454.66,150.27) -- (465.72,150.27) ;
\draw  [dash pattern={on 0.84pt off 2.51pt}]  (446.24,164.65) -- (474.76,164.65) ;

\draw (316,127.5) node [anchor=north west][inner sep=0.75pt]   [align=left] {T-duality?};
\draw (205.67,189.33) node [anchor=north west][inner sep=0.75pt]   [align=left] {Type IIA};
\draw (433,189) node [anchor=north west][inner sep=0.75pt]   [align=left] {Type IIB};

\end{tikzpicture}
\end{align}
And we expect the conjecture can be verified by using the O-vertex/anti-O-vertex method.

\paragraph{Intersecting Defects}
In \cite{Kimura:2021ngu,Nieri:2021xpe}, the 3d supergroup gauge theories arise from intersecting defects in 5d ordinary gauge theories. See also~\cite{Gomis:2016ljm,Pan:2016fbl,Gorsky:2017hro}. The stringy/M-theoretic relation between this two for the orthosymplectic case is of great interest.

\paragraph{AGT Correspondence}
The AGT correspondence means the partition function of 4d $\mathcal{N}=2$ theory with gauge group $G$ on $\mathbb{R}_{\epsilon_{1,2}}^4$ is identified with a conformal block of $W(G)$-algebra \cite{Alday:2009aq,Wyllard:2009hg}. The 5d lift up of this correspondence connects the Nekrasov partition functions of 5d $\mathcal{N} = 1$ gauge theories and symmetries of quantum algebras \cite{Awata:2009ur,Awata:2010yy,Awata:2015hax,Awata:2011ce}. In the sense of BPS/CFT correspondence, we believe there will be underlying algebraic structures in general supergroup gauge theory. Progress has been made in \cite{Noshita:2022dxv} for unitary supergroups, and we expect similiar analysis can be done for the symplectic case also.

\paragraph{Gaiotto Curve} 
The $D_{2|1;\alpha}$ theory which contains three copies of $\mathfrak{sl}(2)$ and tri-fundamental matters resembles the theory we saw in \cite{Gaiotto:2009we}. Its relation with the AGT correspondence can be seen from \cite{Hollands:2010xa,Hollands:2011zc} since we discovered that the partition function has the same structure with the double SO-Sp half-bifundamental formula in \cite[eq.~(A.40)]{Hollands:2010xa}. Therefore we can apply similar analysis to study the $D_{2|1;\alpha}$ supergroup gauge theory.

\paragraph{Quiver Varieties}
The construction of super-instantons suggests a generalization of quiver variety \cite{Nakajima:1994nid,nakajima1998quiver} equipped with supergroup structure \cite{Thind:2010,Bovdi:2020IJAC,Rimanyi:2021hzq}. The exact construction of such quiver varieties in both unitary and orthosymplectic case is worth studying.

\paragraph{Orthosymplectic Superquivers} 
It is also interesting to study orthosymplectic \textit{superquiver} gauge theories with fermionic nodes in the sense of \cite[sec.~8]{Noshita:2022dxv} via $S$-duality as well as its mathematical meaning. Superquiver gauge theories have been realized physically in the sense of super spin chains \cite{Orlando:2010uu,Nekrasov:2018gne,Zenkevich:2018fzl,Ishtiaque:2021jan}. By the philosophy of Bethe/Gauge correspondence, we expect the corresponding gauge theory to be a superquiver gauge theory.

\medskip
These topics will be left for future study.

\appendix
\section*{Acknowledgements}

Y.S. thanks Satoshi Nawata for explaining the effect of fermionic zero modes in orthosymplectic quiver gauge theories and Go Noshita for explaining details on negative branes and T--duality, as well as giving their constructive comments on the manuscript. Y.S. also thanks Cumrun Vafa for his encouragement and beneficial discussions during his stay in ICTP.
Part of this work has been presented in \href{https://indico.in2p3.fr/event/28620/}{French Strings Meeting 2023} at LAPTh, Annecy, France and \href{https://indico.ictp.it/event/10193}{New Pathways in Exploration of Quantum Field Theory and Quantum Gravity beyond Supersymmetry III} at ICTP, Trieste, Italy. Y.S. would like to thank the organizers for their hospitality.
The authors would like to thank the anonymous referees for their valuable advice.
This work was in part supported by EIPHI Graduate School (No.~ANR-17-EURE-0002) and Bourgogne-Franche-Comté region.

\section*{Data Availability}
The work has no associated data.

\section*{Conflict of Interest Statement}
On behalf of all authors, the corresponding author states that there is no conflict of interest. 

\section{Supermathematics}
\subsection{Super Linear Algebra}
We will give a very brief introduction on supergroups first, following \cite{Kimura:2020jxl}. For more about supergroups and its mathematical meanings, we refer the readers to \cite{Frappat:1996pb,DeWitt:2012mdz,varadarajan2004supersymmetry,Berezin:1987wh}.

A \textit{supervector space} is a $\vvmathbb{Z}_2$-graded vector space consists of both commuting and anti-commuting variables
\begin{equation}
    V=V_0 \oplus V_1.
\end{equation}
Its \textit{superdimension}
\begin{equation}
\operatorname{sdim} V=\sum_{\sigma=0,1}(-1)^\sigma \operatorname{dim} V_\sigma=\operatorname{dim} V_0-\operatorname{dim} V_1
\end{equation}
can be negative.

A linear map $M$ between supervector space $V$ and $W$ can be written in terms of a \textit{supermatrix}
\begin{equation}
M=\left(\begin{array}{ll}
M_{00} & M_{01} \\
M_{10} & M_{11}
\end{array}\right)
\end{equation}
with $M_{\sigma \sigma^{\prime}} \in \operatorname{Hom}\left(V_\sigma, W_{\sigma^{\prime}}\right)$. The corresponding \textit{superdeterminant} (or \textit{Berezinian}) is
\begin{equation}
\operatorname{sdet} M=\frac{\operatorname{det}\left(M_{00}-M_{10} M_{11}^{-1} M_{01}\right)}{\operatorname{det} M_{11}}.
\end{equation}
And the \textit{supertrace} is 
\begin{equation}
\operatorname{str} M=\operatorname{tr}_0 M-\operatorname{tr}_1 M=\operatorname{tr} M_{00}-\operatorname{tr} M_{11}.
\end{equation}

Also, we define the \textit{supertranspositon} of a super matrix by
\begin{equation}
M^{\text {st }}=\left(\begin{array}{ll}
M_{00} & M_{01} \\
M_{10} & M_{11}
\end{array}\right)^{\text {st }}=\left(\begin{array}{cc}
M_{00}^{\mathrm{t}} & M_{01}^{\mathrm{t}} \\
-M_{10}^{\mathrm{t}} & M_{11}^{\mathrm{t}}
\end{array}\right)
\end{equation}
where $A^{\mathrm{t}}$ denotes the ordinary transpose operation. We can check that $\left(M_1 M_2\right)^{\text {st }}=M_2^{\text {st }} M_1^{\text {st }}$, but $\left(M^{\text {st }}\right)^{\text {st }} \neq M$ in general. 

The Hermitian conjugation of a supermatrix is defined as follows,
\begin{equation}
M^{\dagger}=\bar{M}^{\mathrm{st}}
\end{equation}
which satisfies $\left(M^{\dagger}\right)^{\dagger}=M$.

\subsection{Supergroups and Superalgebras}
\subsubsection{Classification of Classical Supergroups}
Like the ordinary classification of semi-simple Lie groups (algebras), we can also classify classical supergroups (superalgebras). The results are roughly given by the following table \cite{Quella:2013oda}:
\begin{equation}
\begin{array}{ccc}
\toprule \text{Name} & \text{Alternative name} & \text {Bosonic subgroup} \\
\midrule A_{n|m} & \mathfrak{sl}(n+1 | m+1) & A_n \oplus A_m \oplus T_1 \\
A_{n|n} & \mathfrak{psl}(n | n) & A_n \oplus A_n \\
C_n & \mathfrak{osp}(2 | 2 n-2) & C_{n-1} \oplus T_1 \\
\midrule F_4 & - & A_1 \oplus B_3 \\
G_3 & - & A_1 \oplus G_2 \\
B_{n|m} & \mathfrak{osp}(2 n+1 | m) & B_n \oplus C_m \\
D_{n|m} & \mathfrak{osp}(2 n | m) & D_n \oplus C_m \\
D_{2|1 ; \alpha} & - & A_1 \oplus A_1 \oplus A_1 \\
\toprule
\end{array}
\end{equation}
For details from the mathematical side see \cite{Frappat:1996pb}.

\subsubsection{Unitary Supergroup}
\label{sec:uni_sg}
We define the squared norm 
\begin{equation}
|\Psi|^2=\operatorname{str}\left(\Psi \Psi^{\dagger}\right) .
\end{equation}
for $\Psi \in \vvmathbb{C}^{n|m}$. Then the $A$-type (\textit{unitary supergroup}) $\mathrm{U}(n|m)$ is defined as the isometry group with respect to the supervector space $\vvmathbb{C}^{n|m}$ such that\
\begin{equation}
|\Psi|^2=|U \Psi|^2, \quad U^{\dagger}=U^{-1}
\end{equation}
Thus we have
\begin{equation}
\mathrm{U}(n|m)=\left\{U \in \mathrm{GL}\left(\vvmathbb{C}^{n|m}\right) \mid U^{\dagger}=U^{-1}\right\}
\end{equation}
The bosonic subgroup of $\mathrm{U}(n|m)$ is given by $\mathrm{U}(n) \times \mathrm{U}(m)$, and the fermonic part is given by $\mathbf{n} \times \bar{\mathbf{m}}$ and $\bar{\mathbf{n}} \times \mathbf{m}$ bifundamental representations with the dimension $2nm$.

\subsubsection{Orthosymplectic Supergroup}
\label{sec:osp_sg}
Consider the real supervector space $\vvmathbb{R}^{n|2m}$. The squared norm is defined by:
\begin{equation}
|\Psi|^2=\Psi^{\mathrm{t}} \Omega \Psi=\operatorname{str}\left(\Omega \Psi \Psi^{\mathrm{t}}\right)
\end{equation}
with the skew-symmetric form
\begin{equation}
\Omega=\left(\begin{array}{cc}
\vvmathbb{1}_n & 0 \\
0 & \vvmathbb{1}_m \otimes \mathbf{j}
\end{array}\right), \quad \mathbf{j}=\left(\begin{array}{cc}
0 & 1 \\
-1 & 0
\end{array}\right)
\end{equation}
Then the isometry group of $\vvmathbb{R}^{n|2m}$ is given by
\begin{equation}
|\Psi|^2=|O \Psi|^2, \quad O^{\text {st }} \Omega O=\Omega .
\end{equation}
Thus we have
\begin{equation}
\operatorname{OSp}(n|m)=\left\{O \in \mathrm{GL}\left(\vvmathbb{R}^{n|2m}\right) \mid O^{\mathrm{st}} \Omega O=\Omega\right\}
\end{equation}
The bosonic part of $\OSp(n|m)$ is thus given by $\mathrm{O}(n) \times \mathrm{Sp}(m)$, and the fermonic part is given by $\mathbf{n} \times \mathbf{2m}$ bifundamental representation with the dimension $2nm$.

\subsubsection{Superalgebras}
Superalgebra is also graded
\begin{equation}
\mathfrak{A}=\mathfrak{A}_0 \oplus \mathfrak{A}_1
\end{equation}
with the supercommutator
\begin{equation}
[a, b]=a b-(-1)^{|a||b|} b a
\end{equation}
for $a,b\in\mathfrak{A}$. Here we denote the parity of an element $x \in V_\sigma$ by $|x|=\sigma$ for $\sigma=0,1$, which is called even/bosonic for $\sigma=0$, and odd/fermionic for $\sigma=1$. 
The supercommutator also satisfies the super version of Jacobi identity
\begin{equation}
[a,[b, c]]=[[a, b], c]+(-1)^{|a||b|}[b,[a, c]] .
\end{equation}

For example we denote 
\begin{equation}
\mathfrak{s l}_{n | m}=\left\{a \in \mathfrak{A}=\mathfrak{A}_0 \oplus \mathfrak{A}_1, \operatorname{dim} \mathfrak{A}_0=n, \operatorname{dim} \mathfrak{A}_1=m | \operatorname{str} a=0\right\},
\end{equation}
which corresponds to the $A$-type supergroup $A_{n-1|m-1}$. 

\section{Orientifold Planes}
\label{app:op}
In order to realize the Hanany-Witten construction of $BCD$-type gauge theory, we need to introduce the \textit{orientifold planes} (or O-plane), see \cite{Evans:1997hk,Giveon:1998sr}. A very pedagogical review is \cite[Problem~15.4]{Zwiebach:2004tj}. Here we only briefly introduce the effect of O4-planes.

For a $BCD$-type gauge theory, the brane configuration is:
\begin{align}
\tikzset{every picture/.style={line width=0.75pt}} 
\begin{tikzpicture}[x=0.75pt,y=0.75pt,yscale=-1,xscale=1,baseline=(current bounding box.center)]

\draw    (277.54,84.73) -- (277.6,194.8) ;
\draw [color={rgb, 255:red, 245; green, 166; blue, 35 }  ,draw opacity=1 ] [dash pattern={on 4.5pt off 4.5pt}]  (260.5,194.5) -- (408.5,194.5) ;
\draw    (387.71,111.37) -- (277.64,111.43) ;
\draw    (387.61,84.67) -- (387.67,194.74) ;
\draw    (387.21,169.47) -- (277.14,169.53) ;
\draw    (187.89,161.8) -- (187.7,191.2) ;
\draw [shift={(187.9,159.8)}, rotate = 90.36] [color={rgb, 255:red, 0; green, 0; blue, 0 }  ][line width=0.75]    (10.93,-3.29) .. controls (6.95,-1.4) and (3.31,-0.3) .. (0,0) .. controls (3.31,0.3) and (6.95,1.4) .. (10.93,3.29)   ;
\draw    (187.7,191.2) -- (215.3,191.2) ;
\draw [shift={(217.3,191.2)}, rotate = 180] [color={rgb, 255:red, 0; green, 0; blue, 0 }  ][line width=0.75]    (10.93,-3.29) .. controls (6.95,-1.4) and (3.31,-0.3) .. (0,0) .. controls (3.31,0.3) and (6.95,1.4) .. (10.93,3.29)   ;
\draw    (187.7,191.2) -- (212.31,165.84) ;
\draw [shift={(213.7,164.4)}, rotate = 134.13] [color={rgb, 255:red, 0; green, 0; blue, 0 }  ][line width=0.75]    (10.93,-3.29) .. controls (6.95,-1.4) and (3.31,-0.3) .. (0,0) .. controls (3.31,0.3) and (6.95,1.4) .. (10.93,3.29)   ;

\draw (263.13,65.87) node [anchor=north west][inner sep=0.75pt]   [align=left] {NS5};
\draw (317.8,201.4) node [anchor=north west][inner sep=0.75pt]   [align=left] {O4$\displaystyle ^{\pm }$};
\draw (409.23,131.53) node [anchor=north west][inner sep=0.75pt]   [align=left] {$\displaystyle n$ D4$ $};
\draw (334.23,133.27) node [anchor=north west][inner sep=0.75pt]  [rotate=-90] [align=left] {...};
\draw (373.53,65.07) node [anchor=north west][inner sep=0.75pt]   [align=left] {NS5};
\draw (178.7,143) node [anchor=north west][inner sep=0.75pt]    {$45$};
\draw (218.7,182.6) node [anchor=north west][inner sep=0.75pt]    {$6$};
\draw (207.5,147.8) node [anchor=north west][inner sep=0.75pt]    {$789$};

\end{tikzpicture}
\end{align}
The world-volume of an O-plane reflects the supersymmetry generators, and thus results in the space-time reflection $\left(x^4, x^5, x^7, x^8, x^9\right) \rightarrow\left(-x^4,-x^5,-x^7,-x^8,-x^9\right)$ and the gauging of world sheet parity $\Omega$.

Without the O4-plane, the worldvolume of coincident D4 branes will make up a $\mathrm{U}(n)$ gauge theory. If all D4-branes are coincident and lying on the O4-plane, the open string modes between the D4-branes and their mirror image under O4-plane will give rise to an $\SO\left(2n \right)$ or an $\operatorname{Sp}\left(n\right)$ gauge theory, depending on the choice of whether $\Omega^2=\pm 1$ \cite{Gimon:1996rq}. And for the case of $\SO\left(2n+1\right)$, an extra (half) D4-brane should be placed at on the O4-plane. 

The brane contents of a $BCD$-type gauge theory are listed as follows:
\begin{equation}
\begin{array}{ccccccccccccc}
\toprule \text { Type } & \# & x^0 & x^1 & x^2 & x^3 & x^4 & x^5 & x^6 & x^7 & x^8 & x^9 \\
\midrule \text { NS } & 2 & - & - & - & - & - & - & \bullet & \bullet & \bullet & \bullet \\
\text { O4 } & 1 & - & - & - & - & \bullet & \bullet & - & \bullet & \bullet & \bullet \\
\text { D4 } & n(+1) & - & - & - & - & \bullet & \bullet & |-| & \bullet & \bullet & \bullet \\
\text { D6 } & n^{\rm f}/2 & - & - & - & - & \bullet & \bullet & \bullet & - & - & - \\
\toprule
\end{array}
\end{equation}
in which `$-$' means extending direction and `$\bullet$' means the brane is pointlike in this direction. Notice the D4$^+$ brane is finite along $x^6$ direction. In our $BCD$-type supergroup gauge theory, we also need to consider D4$^-$ branes in addition to ordinary D4$^+$-branes. The extra brane contents are:
\begin{equation}
\begin{array}{ccccccccccccc}
\toprule \text { Type } & \# & x^0 & x^1 & x^2 & x^3 & x^4 & x^5 & x^6 & x^7 & x^8 & x^9 \\
\midrule
\text { D4$^{\pm}$ } & n(+1) & - & - & - & - & \bullet & \bullet & |-| & \bullet & \bullet & \bullet \\
\toprule
\end{array}
\end{equation}

\bibliographystyle{amsalpha_mod}
\bibliography{ref}

\end{document}